\documentclass[12pt]{article}
\usepackage{amssymb}
\usepackage{graphics}
\usepackage{epsfig}
\usepackage{a4wide}

\begin{document}
\newcommand{\eettz}{$e^+e^- \to t\bar tZ^0~$ }
\newcommand{\ggttz}{$\gamma\gamma \to t\bar tZ^0~$ }
\newcommand{\ppttga}{$pp(p\bar p) \to t\bar t\gamma+X~$ }
\newcommand{\ppttgag}{$pp(p\bar p) \to  t\bar t\gamma g+X~$ }
\newcommand{\qqttga}{$q\bar q \to  t\bar t\gamma~$}
\newcommand{\qqttgag}{$q\bar q \to  t\bar t\gamma g~$}
\newcommand{\ggttga}{$gg \to  t\bar t\gamma~$}
\newcommand{\ggttgag}{$gg \to  t\bar t\gamma g~$}
\newcommand{\qgttgaq}{$q(\bar q)g \to  t\bar t\gamma q(\bar q)~$}
\newcommand{\ppuuttga}{$pp(p\bar p) \to u\bar u \to  t\bar t\gamma+X~$}
\newcommand{\ppddttga}{$pp(p\bar p) \to d\bar d \to  t\bar t\gamma+X~$}
\newcommand{\ppqqttga}{$pp(p\bar p) \to q\bar q \to  t\bar t\gamma+X~$}
\newcommand{\ppggttga}{$pp(p\bar p) \to gg \to  t\bar t\gamma+X~$}
\newcommand{\ppgqttga}{$pp(p\bar p) \to gq(g\bar q) \to  t\bar t\gamma q(\bar q)+X~$}

\title{ QCD corrections to associated production of $t\bar t\gamma$ at hadron colliders  }
\author{ Duan Peng-Fei, Ma Wen-Gan, Zhang Ren-You,   \\
Han Liang, Guo Lei, and Wang Shao-Ming  \\
{\small Department of Modern Physics, University of Science and Technology}  \\
{\small of China (USTC), Hefei, Anhui 230026, P.R.China}  }

\date{}
\maketitle \vskip 15mm
\begin{abstract}
We report on the next-to-leading order(NLO) QCD computation of
top-quark pair production in association with a photon at the
Fermilab Tevatron RUN II and CERN Large Hadron Collider. We describe
the impact of the complete NLO QCD radiative corrections to this
process, and provide the predictions of the leading order(LO) and
NLO integrated cross sections, distributions of the transverse
momenta of the top quark and photon for the LHC and Tevatron, and
the LO and NLO forward-backward top-quark charge asymmetries for the
Tevatron. We investigate the dependence of the LO and NLO cross
sections on the renormalization/factorization scale, and find the
scale dependence of the LO cross section is obviously improved by
the NLO QCD corrections. The K-factor of the NLO QCD correction is
$0.977(1.524)$ for the Tevatron(LHC).
\end{abstract}


\vfill \eject

\baselineskip=0.32in

\renewcommand{\theequation}{\arabic{section}.\arabic{equation}}
\renewcommand{\thesection}{\Roman{section}.}
\newcommand{\nb}{\nonumber}

%slash:
\newcommand{\Dir}{\kern -6.4pt\Big{/}}%su lettere italiane minuscole
\newcommand{\Dirin}{\kern -10.4pt\Big{/}\kern 4.4pt}
\newcommand{\DDir}{\kern -7.6pt\Big{/}}%su lettere italiane maiuscole
\newcommand{\DGir}{\kern -6.0pt\Big{/}}%su lettere greche

\makeatletter      % '@' is now a normal "letter" for TeX
\@addtoreset{equation}{section}
\makeatother       % '@' is restored as a "non-letter" character for TeX

\section{Introduction}
\par
The top-quark was discovered by the CDF and D0 collaborations at
Fermilab Tevatron in 1995\cite{cdftop,d0top}. It opens up a new
research field of top physics, and confirms again the
three-generation structure of the standard model(SM)\cite{s1,s2}.
Among all the elementary particles discovered up to
now\cite{tew,hepdata}, the top-quark mass term breaks the
electroweak (EW) gauge symmetry maximally due to its huge mass, and
the detailed physics of the top-quark may be significantly different
from the predictions provided by the SM. But until now our knowledge
about top quark's properties has been still limited \cite{Chak}. For
example, the couplings of the top quark to a photon and a $Z^0$
boson have not yet been directly measured\cite{Smith,ttz}, while the
precise measurement of the production and decay of top quark may be
significant in searching for new physics beyond the SM.

\par
In recent years there have been many works devoted to the study of
the top-quark couplings. The studies for probing the top-quark
couplings $t\bar t\gamma$ and $t\bar tZ^0$ at hadron colliders at LO
were carried out in Ref.\cite{UBaur}, the calculations for the
process \eettz at LO, QCD and EW NLO are provided in
Refs.\cite{Hagiwara}, and the one-loop SM QCD and the supersymmetric
QCD effects in the process of \ggttz at the ILC was investigated in
Ref.\cite{Dong}. The SM couplings of $t\bar tV(V=\gamma,Z^0)$ may be
modified by the new interactions and that would lead to abundant
phenomena of new physics. For example, if the top quark was a
composite object, there would be an anomalously large $t\bar
t\gamma$ event rate at colliders, due to deexcitation of
high-energetic top state\cite{Werner}. And if there exists
nonstandard CP violation, in particular Higgs sector CP violation, a
sizable top-quark (weak) electric dipole moment could be
induced\cite{pham}. Other relevant
references\cite{ttz,examples,Hill,Berger} indicate that the vector
and axial form factors in the coupling of the top quark and neutral
gauge boson $V(=\gamma, Z^0)$ should be probed precisely in order to
find the signatures of a certain model of dynamical EW breaking.

\par
At a linear collider, it is not easy to obtain the information about
the individual EW neutral coupling $t\bar tV$ $(V=Z^0, \gamma)$ from
the precise measurement of the top-pair production at a linear
collider because of the hardness in distinguishing the contributions
from the $t\bar tZ^0$ and $t\bar t\gamma$ couplings. At a hadron
collider, it is impossible to measure the EW neutral couplings via
$q\bar{q}\to\gamma^*/Z^*\to t\bar{t}$ due to the strong interaction
process $q\bar{q}(gg)\to g^*\to t\bar{t}$. Instead, they can be
measured in QCD $t\bar{t}Z^0/\gamma$ production and radiative
top-quark decays in $t\bar{t}$ events ($t\bar{t}\to\gamma W^+W^-
b\bar{b}$). Each of the processes is sensitive to the EW coupling
between the top quark and the emitted $Z^0$-boson(or photon). In the
work of \cite{Baur} it is concluded that it will be possible to
probe the $t\bar t\gamma$ coupling at a few percent level at the
LHC. Since the LO predictions in the QCD expansion for the channels
$pp(p\bar p) \to t\bar tZ^0(\gamma)+X$ at hadron colliders contain
significant theoretical uncertainty, it is important to improve the
theoretical prediction in order to accommodate the experimental
measurement of the top-quark couplings. Recently, the NLO QCD
correction to $t\bar t Z^0$ production at the LHC has been
calculated in Ref.\cite{pp-ttz}.

\par
Our study in this work corresponds to the investigation on the
production of the top-quark pair associated with a photon at the
Fermilab Tevatron Run II and the CERN LHC in both LO and NLO QCD
approximations. It is arranged as follows: In Sec.II we provide
descriptions of the analytical calculations. In Sec.III we present
some numerical results and discussions, and finally a short summary
is given.

\par
\section{Description of the calculation   }
\par
In the calculations at the LO and NLO of the $\alpha_s$ expansion,
we use the 't Hooft-Feynman gauge, employ FeynArts3.4
package\cite{fey} to generate Feynman diagrams and their
corresponding amplitudes. The LO amplitudes are precessed by
adopting FormCalc5.4 programs\cite{formloop}. In the calculation for
virtual corrections, the one-loop amplitudes involving UV and IR
singularities are handled analytically by using our modified
FormCalc programs, and are output in Fortran code with the UV and IR
``$\epsilon \times$ N-point integrals'' terms remained unprocessed.
The output is further processed numerically by using our developed
Fortran subroutines for calculating N-point integrals to extract the
remaining finite $\epsilon \frac{1}{\epsilon}$ terms. In these
Fortran codes the IR singularities are separated from the IR-finite
remainder by adopting the expressions for the IR singularity in
N-point integrals($N\geq3$) in terms of 3-point
integrals\cite{Beenakk}.

\par
\subsection{Born approximation  }
\par
\begin{figure*}
\begin{center}
\includegraphics[scale=1.0]{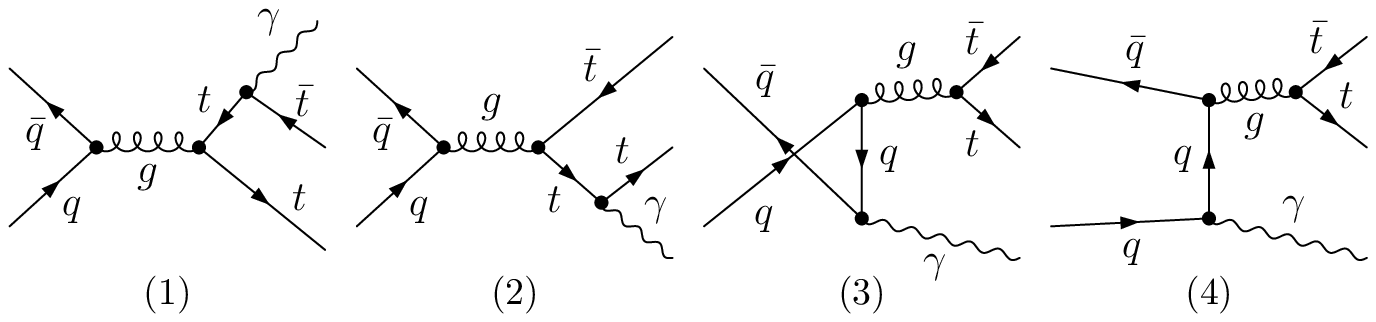}
\caption{\label{fig1a} The LO Feynman diagrams for the
\qqttga($q=u,d,s,c$) partonic process. }
\end{center}
\end{figure*}
\begin{figure*}
\begin{center}
\includegraphics[scale=1.0]{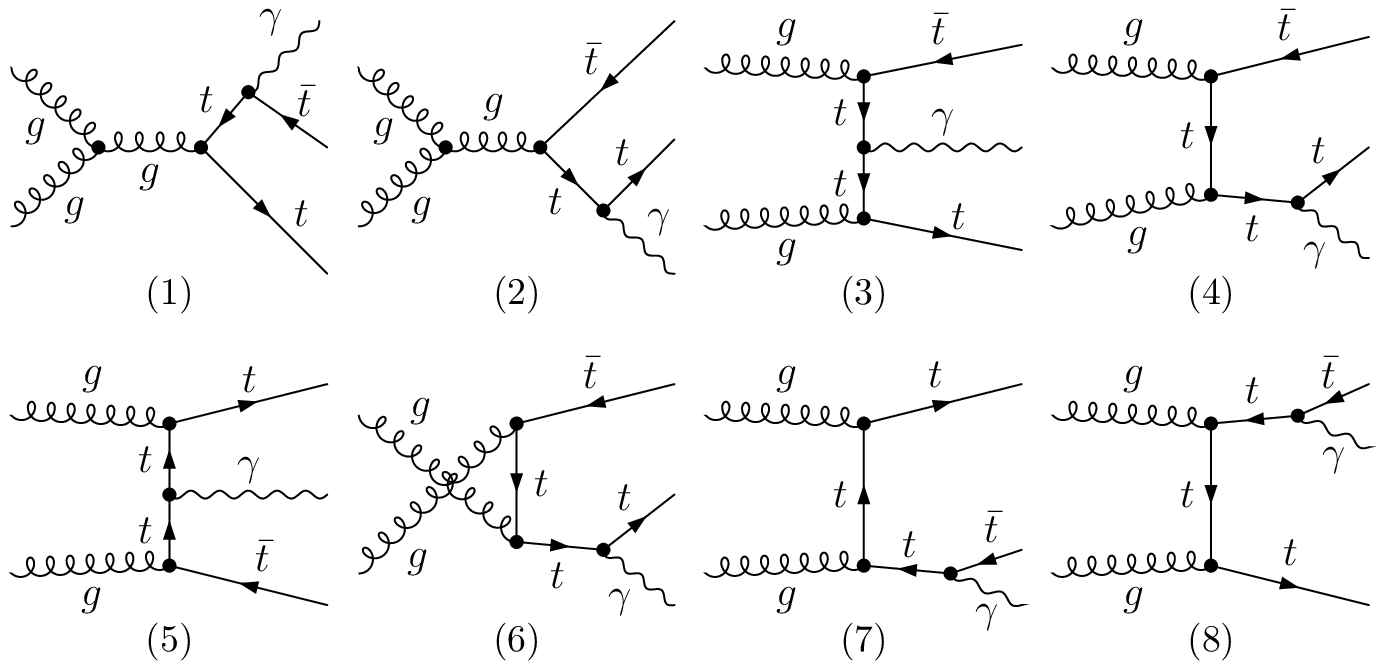}
\caption{\label{fig1b} The LO Feynman diagrams for the \ggttga
partonic process. }
\end{center}
\end{figure*}

We consider five partonic processes which contribute to the process
of top-pair production associated with a photon at LO for hadron
colliders. They are \ggttga and \qqttga($q=u,d,c,s$) production
channels. We take the constraint for the transverse momentum for
radiated photon as $p_T^{(\gamma)} > p_{T,cut}^{(\gamma)}$, e.g.,
$p_{T,cut}^{(\gamma)}=20~GeV$.  We express these partonic reactions
as
\begin{equation}
\label{process1} q(p_1)+\bar q(p_2) \to t(p_3)+ \bar t
(p_4)+\gamma(p_5),~~~~(q=u,d,s,c),
\end{equation}
and
\begin{equation}
\label{process2} g(p_1)+g(p_2) \to t(p_3)+ \bar t (p_4)+\gamma(p_5),
\end{equation}
where we denote the external four-momenta by ~${p_i}$$(i=1,...,5)$
for the partonic processes \qqttga and \ggttga, separately. There
are 4 LO Feynman diagrams for partonic process \qqttga(shown in
Fig.\ref{fig1a}), and 8 tree-level diagrams for the \ggttga partonic
process(shown in Fig.\ref{fig1b}). Despite being massless for photon
and light-quarks($q=u,d,s$), the cross sections for the above
partonic processes are still "infrared safe" due to our constraint
for the photon transverse momentum.
%%figure%%
%%figure%%
%%figure%%

\par
The expression of LO cross section for the partonic processes
\qqttga and \ggttga have the forms respectively as
\begin{eqnarray}\label{sigma_qqgg}
d\hat{\sigma}^{0}_{q\bar q}= \frac{1}{4}\frac{1}{9}\frac{(2 \pi
)^4}{2\hat{s}} \sum_{spin}^{color} |{\cal M}_{LO}^{q\bar q}|^2
d\Omega_{3}^{q\bar q},~~~d\hat{\sigma}^{0}_{gg}=
\frac{1}{4}\frac{1}{64}\frac{(2 \pi )^4}{2\hat{s}}
\sum_{spin}^{color} |{\cal M}_{LO}^{gg}|^2 d\Omega_{3}^{gg},
\end{eqnarray}
where the factors $\frac{1}{4}$, $\frac{1}{9}$ and factors
$\frac{1}{4}$, $\frac{1}{64}$ in Eqs.(\ref{sigma_qqgg}) come from
the averaging over the spins and colors of the initial partons,
respectively, $\hat{s}$ is the partonic center-of-mass energy
squared, ${\cal M}_{LO}^{q\bar q}$ and ${\cal M}_{LO}^{gg}$ are the
amplitudes of all the tree-level diagrams for the partonic processes
\qqttga and \ggttga respectively. In above two equations the
summations are taken over the spins and colors of all the relevant
particles in the \qqttga and \ggttga partonic processes. The
phase-space elements $d\Omega_{3}^{q\bar q}$ and $d\Omega_{3}^{gg}$
in Eqs.(\ref{sigma_qqgg}) is expressed as
\begin{eqnarray}\label{PhaseSpace}
{d\Omega_{3}^{q\bar q,gg}}=\delta^{(4)} \left( p_1+p_2-\sum_{i=3}^5
p_i \right) \prod_{j=3}^5 \frac{d^3 \textbf{\textsl{p}}_j}{(2 \pi)^3
2 E_j}.
\end{eqnarray}

\par
According to the factorization theorem for hard scattering processes
in QCD, the LO differential cross section for the process $p\bar
p(pp) \to t\bar t \gamma+X$ at the Tevatron(LHC) can be obtained by
performing the following integration of the differential cross
section for the partonic processes \qqttga and \ggttga over the
initial partonic luminosities [see Eq.(\ref{integration})].
\begin{eqnarray}\label{integration}
d \sigma_{LO} &=& \sum_{ij=u\bar{u},d\bar{d}}^{s\bar{s},c\bar{c},gg}
              \int_{0}^1 dx_1 \int_0^1 dx_2
              \frac{1}{1+\delta_{ij}}  \nonumber \\
&&             \times  \Big[
                  G_{i/P_1} \left( x_1,\mu_f \right) G_{j/P_2} \left( x_2,\mu_f \right)
                  \frac{d \hat{\sigma}^0_{ij}}{d \hat{y}_t}
               + G_{j/P_1} \left( x_1,\mu_f \right) G_{i/P_2} \left( x_2,\mu_f \right)
                  \frac{d \hat{\sigma}^0_{ji}}{d \hat{y}_t}
              \Big] d y_t,
\end{eqnarray}
where $y_t$ and $\hat {y}_t$ are the rapidities of the top-quark in
the proton-(anti)proton and partonic center-of-mass systems,
respectively
($y_t=\frac{1}{2}\ln\left(\frac{E^t+p_z^t}{E^t-p_z^t}\right)$,
$\hat{y}_t=\frac{1}{2}\ln\left(\frac{\hat E^t+\hat{p}_z^t}{\hat
E^t-\hat{p}_z^t}\right)$). The direction of the z-axis of the
hadronic center-of-mass system is defined as the orientation of
incoming hardron $P_1$(for the parent process $P_1 P_2 \to t \bar{t}
\gamma + X$), while the z-axis of the partonic center-of-mass system
is set as the orientation of radiated parton $i(or~j)$ from
$P_1$[for the partonic process $i j(or~ji) \to t \bar{t} \gamma$].
The differential cross sections $\frac{d \hat{\sigma}^0_{ij}}{d
\hat{y}_t}$ and $\frac{d \hat{\sigma}^0_{ji}}{d \hat{y}_t}$ are
expressed in their own partonic center-of-mass frames, respectively.
Under a boost in the z-direction to a frame with velocity $\beta$,
we have $y_t = \hat{y}_t-\tanh^{-1}\beta$ and $d y_t=d \hat{y}_t$.
$G_{i(j)/A}(x,\mu_f)$($i=u,d,s,c,g,~j=\bar u,\bar d,\bar s,\bar
c,g$) are the parton distribution functions(PDFs) of (anti)proton
$A(=P_1,P_2)$ which describe the probability to find a parton $i(j)$
with four-momentum $xp_A$ in (anti)proton $A$. The partonic
colliding energy squared $\hat{s}=x_1x_2 s$, where $s$ is defined as
the center-of-mass energy squared of the proton-(anti)proton
collision. $\mu_f$ is the factorization energy scale. In our LO
calculations, we adopt the CTEQ6L1 PDFs\cite{pdfs}.

\par
Our LO calculation shows when we take $p_{T,cut}^{(\gamma)}=20~GeV$,
the LO integrated cross section for the $t\bar t\gamma$ production
is dominated by the gluon-gluon fusion partonic channel with about
$66.3\%$ at the LHC , while about $99.3\%$ is contributed by the
$q-\bar q(q=u,d)$ annihilation partonic channels at the Tevatron RUN
II.

\par
\subsection{NLO QCD corrections }
\par
The NLO QCD corrections to the \ppttga process are contributed
distinctly by the following four parts:
\par
1. the real gluon emission partonic processes $q \bar{q}, gg \to t
\bar{t} \gamma g,~(q = u,d,s,c)$.
\par
2. the real light-(anti)quark emission partonic processes
$q(\bar{q}) g \to t \bar{t} \gamma q(\bar{q}), ~(q = u,d,s)$.
\par
3. the virtual corrections at the NLO to the partonic processes $q
\bar{q}, gg \to t \bar{t} \gamma, ~(q = u,d,s,c)$.
\par
4. the collinear counterterms of the PDF.

\par
In all the NLO calculations we use the dimensional
regularization(DR) method in $D=4-2 \epsilon$ dimensions to isolate
the UV and IR singularities. To describe the cancelations of the IR
singularities in our calculations more clearly, we decompose the
collinear counterterms of the PDF, $\delta G_{i/P}(x,\mu_f)~
(P=p,\bar{p}~;~i=g, u, \bar{u}, d, \bar{d}, s, \bar{s})$, into two
parts: the collinear gluon emission part $\delta
G_{i/P}^{(gluon)}(x,\mu_f)$ and the collinear light-quark emission
part $\delta G_{i/P}^{(quark)}(x,\mu_f)$. Their analytical
expressions are presented as follows.
\begin{eqnarray}\label{PDFcounterterm1}
&& \delta G_{q(g)/P}(x,\mu_f) = \delta G_{q(g)/P}^{(gluon)}(x,\mu_f)
                            +\delta G_{q(g)/P}^{(quark)}(x,\mu_f),
                            ~~(q = u, \bar{u}, d, \bar{d}, s, \bar{s}),
\end{eqnarray}
where
\begin{eqnarray}\label{PDFcounterterm2}
&& \delta G_{q(g)/P}^{(gluon)}(x,\mu_f) =
   \frac{1}{\epsilon} \left[
                      \frac{\alpha_s}{2 \pi}
                      \frac{\Gamma(1 - \epsilon)}{\Gamma(1 - 2 \epsilon)}
                      \left( \frac{4 \pi \mu_r^2}{\mu_f^2} \right)^{\epsilon}
                      \right]
   \int_z^1 \frac{dz}{z} P_{qq(gg)}(z) G_{q(g)/P}(x/z,\mu_f), \nonumber \\
&& \delta G_{q/P}^{(quark)}(x,\mu_f) =
   \frac{1}{\epsilon} \left[
                      \frac{\alpha_s}{2 \pi}
                      \frac{\Gamma(1 - \epsilon)}{\Gamma(1 - 2 \epsilon)}
                      \left( \frac{4 \pi \mu_r^2}{\mu_f^2} \right)^{\epsilon}
                      \right]
   \int_z^1 \frac{dz}{z} P_{qg}(z) G_{g/P}(x/z,\mu_f),  \nonumber \\
&& \delta G_{g/P}^{(quark)}(x,\mu_f) =
   \frac{1}{\epsilon} \left[
                      \frac{\alpha_s}{2 \pi}
                      \frac{\Gamma(1 - \epsilon)}{\Gamma(1 - 2 \epsilon)}
                      \left( \frac{4 \pi \mu_r^2}{\mu_f^2} \right)^{\epsilon}
                      \right]
   \sum_{q=u,\bar{u}}^{d,\bar{d}, s, \bar {s}}
   \int_z^1 \frac{dz}{z} P_{gq}(z) G_{q/P}(x/z,\mu_f).
\end{eqnarray}

\par
The virtual corrections to the processes $pp(p\bar{p}) \to
q\bar{q},gg \to t\bar{t}\gamma + X$ contain both soft and collinear
IR singularities. These singularities can be canceled exactly by
adding the contributions of the real gluon emission processes
$q\bar{q},gg \to t\bar{t}\gamma g$ and the collinear gluon emission
part of the PDF counterterms $\delta G_{q(g)/P}^{(gluon)}$. The real
light-quark emission processes $q(\bar{q})g \to t\bar{t}\gamma
q(\bar{q})$ contain only the collinear IR singularities. It can be
canceled by the contributions of the collinear light-quark emission
part of the PDF counterterms $\delta G_{q(g)/P}^{(quark)}$ exactly.
All of these cancelations are verified numerically in our numerical
calculations. The explicit expressions for the splitting functions
$P_{ij}(z),~(ij=qq,qg,gq,gg)$ can be found in Ref.\cite{Harris}.

\vskip 5mm
{\bf A. Real gluon emission corrections }
\vskip 5mm

We denote the partonic processes with real gluon emissions as
\begin{equation}
q(p_1)+\bar q (p_2) \to t(p_3) +\bar t(p_4) + \gamma(p_5) +
g(p_6),~~ g(p_1)+g(p_2) \to t(p_3) +\bar t(p_4) + \gamma(p_5) +
g(p_6).
\end{equation}

\par
\begin{figure*}
\begin{center}
\includegraphics[scale=1.0]{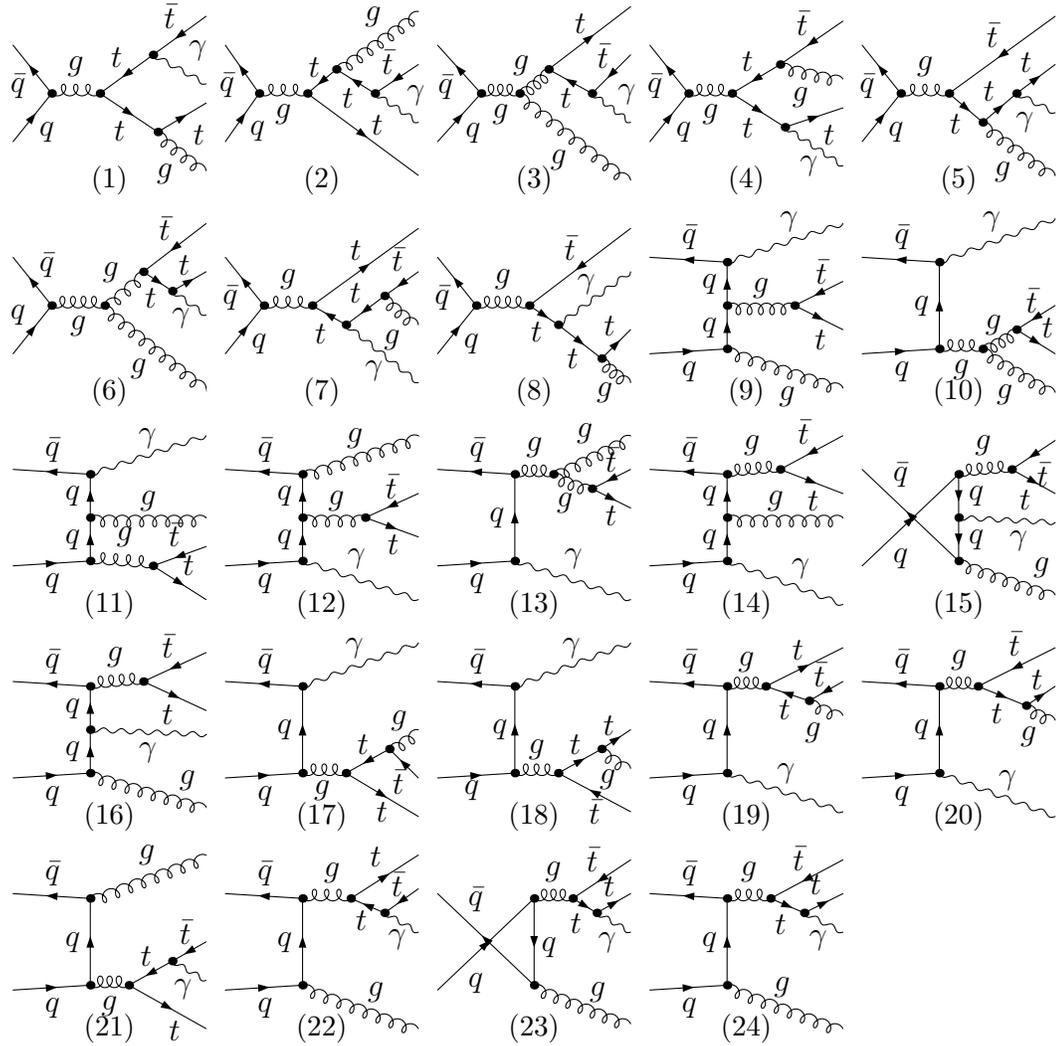}
\vspace*{-0.3cm} \centering \caption{\label{fig2a} The LO Feynman
diagrams for the real gluon emission partonic process
\qqttgag$(q=u,d,s,c)$. }
\end{center}
\end{figure*}
\begin{figure*}
\begin{center}
\includegraphics[scale=1.0]{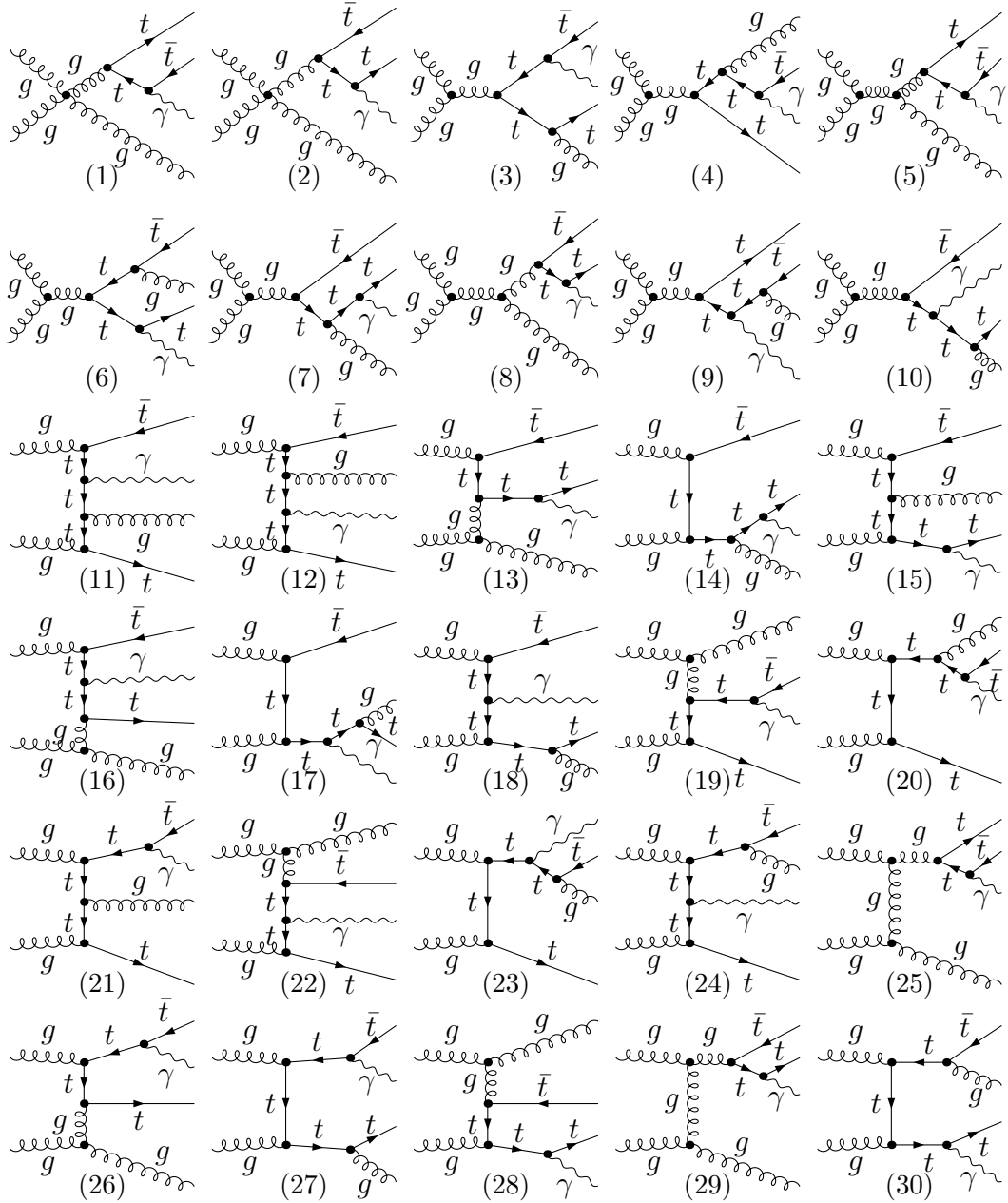}
\vspace*{-0.3cm} \centering \caption{\label{fig2b} The LO Feynman
diagrams for the real gluon emission partonic process \ggttgag. The
diagrams obtained by exchanging two initial gluon lines in (11)-(30)
are not drawn. }
\end{center}
\end{figure*}

The real gluon emission partonic process \qqttgag includes 24 LO
graphs shown in Fig.\ref{fig2a}, and the \ggttgag subprocess
involves 50 LO graphs(shown in Fig.\ref{fig2b}). The figures
(1)-(10) in Fig.\ref{fig2b} are s-channel diagrams, the
Figs.\ref{fig2b}(11)-(30) are t-channel diagrams. The u-channel
diagrams for the \ggttgag subprocess are not drawn in
Fig.\ref{fig2b}, but can be obtained by exchanging incoming gluons
in each t-channel diagram in Fig.\ref{fig2b}. The process $c \bar{c}
\to t \bar{t} \gamma g$ contains only the soft IR singularity, while
$q \bar{q} \to t \bar{t} \gamma g, ~(q = u,d,s)$ and $gg \to t
\bar{t} \gamma g$ contain both soft and IR singularities. The soft
IR singularities can be extracted by adopting the two cutoff
phase-space slicing(TCPSS) methods\cite{Harris} respectively. The
soft IR singularities in the partonic processes $q \bar{q} \to t
\bar{t} \gamma g, ~(q=u,d,s,c)$ and $gg \to t \bar{t} \gamma g$ at
the LO cancel the corresponding soft IR singularities arising from
the one-loop virtual corrections to $q \bar{q} \to t \bar{t} \gamma,
~(q = u,d,s,c)$ and $gg \to t \bar{t} \gamma$ processes,
respectively.
%%figure%%
%%%%figure%%%%

\par
We split the phase-space of the $gg(q \bar{q}) \to t \bar{t} \gamma
g, ~(q = u,d,s,c)$ partonic process into two regions, $E_6 \leq
\delta_s\sqrt{\hat{s}}/2$(soft gluon region) and $E_6
> \delta_s\sqrt{\hat{s}}/2$(hard gluon region). Except for the
$c \bar{c} \to t \bar{t} \gamma g$ process, the hard gluon region is
divided into hard collinear region(${\rm HC}$)($-\hat{t}_{16}$ or
$-\hat{t}_{26}$ $<\delta_c \hat{s}$) and hard noncollinear
($\overline{\rm HC}$) region($-\hat{t}_{16}$ or $-\hat{t}_{26} \geq
\delta_c \hat{s}$), where $\hat{t}_{ij} = (p_i-p_j)^2$ and
$(p_i-p_j)^2$ for the $q \bar{q} \rightarrow t \bar{t} \gamma g$ and
$gg \rightarrow t \bar{t} \gamma g$ respectively. Then the cross
sections for the real gluon emission partonic processes can be
expressed as
\begin{eqnarray}
&& \hat{\sigma}^R_{g,ij} \left( ij \to t \bar{t} \gamma g \right)
 = \hat{\sigma}^S_{g,ij} + \hat{\sigma}^H_{g,ij}~~,~~~~~
(ij = u\bar{u}, d\bar{d}, s\bar{s}, c\bar{c}, gg) \nonumber \\
&& \hat{\sigma}^H_{g,ij} \left( ij \to t \bar{t} \gamma g \right)
= \hat{\sigma}^{HC}_{g,ij} +
\hat{\sigma}^{\overline{HC}}_{g,ij}~~, ~~~~~(ij = u\bar{u},
d\bar{d}, s\bar{s}, gg)
\end{eqnarray}

\par
The differential cross section for the
partonic processes \qqttgag in the soft region is given as
\begin{eqnarray}\label{softIn-1}
d\hat{\sigma}^{S}_{g,q\bar q}&=&-\frac{\alpha_s}{2\pi}\left[
\frac{1}{6}(g_{12}+g_{34})-\frac{7}{6}(g_{13}+g_{24})-\frac{1}{3}(g_{14}+g_{23})\right]
d\hat{\sigma}^0_{q\bar q} \nonumber \\
&=& \left[\frac{\alpha_s}{2\pi}\frac{\Gamma (1-\epsilon)}{\Gamma
(1-2\epsilon)}\left(\frac{4\pi\mu_r}{\hat{s}}
\right)^\epsilon\right]\left(\frac{A_{2,q\bar q}
^S}{\epsilon^2}+\frac{A_{1,q\bar q} ^S}{ \epsilon}+A_{0,q\bar
q}^S\right)d\hat{\sigma}^0_{q\bar q},
\end{eqnarray}
where $d\hat{\sigma}^0_{ij}$ are the LO differential cross
sections for the partonic processes $ij \to t\bar t \gamma$,
$(ij=u\bar{u}, d\bar{d}, s\bar{s}, c\bar{c})$. The soft integrals
$g_{ij}(i=1,2,3,~j=2,3,4)$ are defined as
\begin{eqnarray}\label{eq1}
g_{ij}(p_i,p_j)=\frac{(2\pi\mu_r)^{2\epsilon}}{2\pi}\int_{E_6\leq
\delta_s\sqrt{\hat{s}/2}} \frac{d^{D-1}{\bf
p_6}}{E_6}\left[\frac{2(p_i \cdot p_j)}{(p_i \cdot p_6)(p_j \cdot
p_6)}- \frac{p_i^2}{(p_i \cdot p_j)^2}-\frac{p_j^2}{(p_j \cdot
p_6)^2}\right].
\end{eqnarray}
The explicit expressions for the soft integrals $g_{ij}(p_i,p_j)$
relevant to our calculations for the $q\bar q, gg \to t\bar t \gamma
g$ partonic processes, can be found in Ref.\cite{ppqcd}. By using
Eqs.(\ref{softIn-1}-\ref{eq1}) and the related soft integral
expressions, we can express the coefficients $A_{2,q\bar q}^S$ and
$A_{1,q\bar q}^S$ in Eq.(\ref{softIn-1}) for the massless $q\bar
q$-fusion processes \qqttgag($q=u,d,s$) in the forms as
\begin{eqnarray}
A_{2,q\bar q}^S&=&\frac{8}{3},     \nonumber\\
A_{1,q\bar q}^S&=&\frac{8}{3}-\frac{16}{3}\delta_s-\frac{1}{3}
\frac{p_3 \cdot p_4}{\lambda^{1/2}(s_{34},m_t^2,m_t^2)}
\log\left(\sigma_{34}\sigma_{43}\right)  \nonumber \\
&&-\frac{7}{3}\left(\log \frac{p_1 \cdot p_3}{p_1^0m_t} +\log
\frac{p_2 \cdot p_4}{p_2^0m_t}\right) - \frac{2}{3} \left(\log
\frac{p_2 \cdot p_3}{p_2^0m_t}+\log \frac{p_1 \cdot
p_4}{p_1^0m_t}\right).
\end{eqnarray}
And for the massive $c\bar c$-fusion partonic process $c\bar c \to
t\bar t \gamma g$, we get
\begin{eqnarray}
A_{2,q\bar q}^S&=&0, \nonumber \\
A_{1,q\bar q}^S&=&\frac{16}{3}-\frac{1}{3}\left(\frac{p_1 \cdot
p_2}{\lambda^{1/2}(s_{12},m_c^2,m_c^2)}
\log\left(\sigma_{12}\sigma_{21}\right)+\frac{p_3 \cdot
p_4}{\lambda^{1/2}(s_{34},m_t^2,m_t^2)}
\log\left(\sigma_{34}\sigma_{43}\right)\right) \nonumber \\
&&+\frac{7}{3}\left(\frac{p_1 \cdot
p_3}{\lambda^{1/2}(s_{13},m_c^2,m_t^2)}
\log\left(\sigma_{13}\sigma_{31}\right)+\frac{p_2 \cdot
p_4}{\lambda^{1/2}(s_{24},m_c^2,m_t^2)}
\log\left(\sigma_{24}\sigma_{42}\right)\right) \nonumber \\
&&+\frac{2}{3}\left(\frac{p_1 \cdot
p_4}{\lambda^{1/2}(s_{14},m_c^2,m_t^2)}
\log\left(\sigma_{14}\sigma_{41}\right)+\frac{p_2 \cdot
p_3}{\lambda^{1/2}(s_{23},m_c^2,m_t^2)}
\log\left(\sigma_{23}\sigma_{32}\right)\right).
\end{eqnarray}
with $\sigma_{ij}=\frac{1-\rho_{ij}}{1+\rho_{ij}}$,
$\rho_{ij}=\frac{\lambda^{1/2}(s_{ij},m_i^2,m_j^2)}{
s_{ij}+m_i^2-m_j^2}$ and
$\lambda^{1/2}(s_{ij},m_i^2,m_j^2)=\sqrt{(s_{ij}+m_i^2-m_j^2)^2-
4s_{ij}m_i^2}$.

\par
For the \ggttgag partonic process in the soft region, we have
\begin{eqnarray}\label{eq5}
d\hat{\sigma}^{S}_{g,gg}&=&\frac{\alpha_s}{12\pi}\overline{\sum}\left[\left(\frac{256}{3}D_1+16D_3\right)|
{\cal M}_1^{gg}|^2
+\left(\frac{256}{3}D_2+16D_4\right)|{\cal M}_2^{gg}|^2 \right. \nonumber \\
&&\left. +\left(-\frac{32}{3}D_1+16D_3\right) 2{\bf Re}({\cal
M}_1^{gg\dag} \cdot {\cal M}_2^{gg})\right]d\Omega_3^{gg},
\end{eqnarray}
where the summation is taken over the spins and colors of initial
and final states, and the bar over the summation represents taking
average over the spins and colors of initial partons, and
\begin{eqnarray}\label{eq6}
{\cal M}_1^{gg}={\cal M}_t^{gg}+\frac{1}{2}{\cal M}_s^{gg},~~~
{\cal M}_2^{gg}={\cal M}_u^{gg}-\frac{1}{2}{\cal M}_s^{gg},
\end{eqnarray}
${\cal M}_{s}^{gg}$, ${\cal M}_{t}^{gg}$ and ${\cal M}_{u}^{gg}$
are the amplitudes for $s$-, $t$- and $u$-channel diagrams of
partonic process \ggttga separately, and
\begin{eqnarray}\label{eq7}
{\cal M}_{LO}^{gg}=
\left(\frac{2}{3}C_1^{gg}+C_2^{gg}+C_3^{gg}\right){\cal M}_1^{gg}+
\left(\frac{2}{3}C_1^{gg}-C_2^{gg}+C_3^{gg}\right){\cal M}_2^{gg}.
\end{eqnarray}
The color factors are expressed as
\begin{eqnarray}\label{eq8}
C_1^{gg}=\delta^{c_1c_2}{\bf 1},\quad C_2^{gg}=i
f^{c_1c_2c}\lambda^c,\quad C_3=d^{c_1c_2c}\lambda^c,
\end{eqnarray}
where $f^{abc}$ and $d^{abc}$ are antisymmetric and symmetric
$SU(3)$ structure constants respectively, ${\bf 1}$ and
$\lambda^c$ are identity and Gell-Mann matrices. $c_1,c_2$ are the
color indices of initial gluons, $c$ is the color index of
propagator gluon and
\begin{eqnarray}\label{eq9}
D_1&=&9g_{12}+9g_{13}+9g_{24}-g_{34},~~
D_2=9g_{12}+9g_{23}+9g_{14}-g_{34},\nonumber \\
D_3&=&6(g_{12}-g_{14}-g_{23}+g_{34}),~~
D_4=6(g_{12}-g_{13}-g_{24}+g_{34}).
\end{eqnarray}
The cross sections for the processes $pp(p\bar{p}) \rightarrow ij
\rightarrow t \bar{t} \gamma g + X, ~(ij = u\bar{u}, d\bar{d},
s\bar{s}, gg)$ in the hard noncollinear region,
$\hat{\sigma}^{\overline{HC}}_{g,ij}$, and $pp(p\bar{p})
\rightarrow c \bar{c} \rightarrow t \bar{t} \gamma g + X$ in the
hard region, $\hat{\sigma}^{H}_{g,c\bar{c}}$, are finite and can
be calculated by using Monte Carlo method. The differential cross
section in the hard collinear region, $d\sigma^{HC}_{g,ij}$, can
be obtained by using
\begin{eqnarray}\label{collinear-g}
 d \sigma^{HC}_{g,ij}
 &=&
    \frac{1}{1+\delta_{ij}}
    \left[ \frac{\alpha_s}{2 \pi}
           \frac{\Gamma \left( 1 - \epsilon \right)}{\Gamma \left( 1 - 2 \epsilon \right)}
           \left( \frac{4 \pi \mu_r^2}{\hat{s}}
           \right)^{\epsilon}
    \right]
    \left( -\frac{1}{\epsilon} \right) \delta_c^{-\epsilon}
    \int_0^1 dx_1 \int_0^1 dx_2  \nonumber \\
 && \Big\{
    \Big[
    \int_{x_1}^{1-\delta_s} \frac{dz}{z} \left( \frac{1-z}{z} \right)^{-\epsilon}
    P_{ii} \left( z,\epsilon \right)
    G_{i/P_1} \left( x_1/z, \mu_f \right) G_{j/P_2} \left( x_2, \mu_f \right) \nonumber \\
&&  +~
    \int_{x_2}^{1-\delta_s} \frac{dz}{z} \left( \frac{1-z}{z} \right)^{-\epsilon}
    P_{jj} \left( z,\epsilon \right)
    G_{i/P_1} \left( x_1, \mu_f \right) G_{j/P_2} \left( x_2/z, \mu_f \right)
    \Big]
    d \hat{\sigma}^0_{ij} \nonumber \\
&&+ ~\left( i \leftrightarrow j \right)
    \Big\},
\end{eqnarray}
where $G_{i(j)/P}(x, \mu_f)$ is the PDF of parton $i(j)$, and P
refers to (anti)proton. $P_{ii}(z,\epsilon)$ $(i=q$ for $q-\bar q$
annihilation subprocess and $i=g$ for $g-g$ fusion subprocess) are
the $D$-dimensional unregulated ($z<1$) splitting functions related
to the usual Altarelli-Parisi splitting kernel \cite{Altarelli}.
They can be written explicitly as
\begin{eqnarray}
P_{ii}(z,\epsilon)=P_{ii}(z)+ \epsilon P'_{ii}(z),~~~(i=q,g), \nb
\end{eqnarray}
\begin{eqnarray}
P_{qq}(z)=C_F \frac{1+z^2}{1-z},~~~ P'_{qq}(z)=-C_F (1-z), \nb \\
P_{gg}(z)=2N\left [ \frac{z}{1-z}+\frac{1-z}{z}+z(1-z) \right ],~~~
P'_{gg}(z)=0,
\end{eqnarray}
where $N=3$ is the color number, $C_F=4/3$.

\vskip 5mm
{\bf B. Real light-(anti)quark emission corrections  }
\vskip 5mm

Since the LO contributions from the real light-(anti)quark
emission partonic processes \qgttgaq are at the same $\alpha_s$
order as previous real gluon emission partonic processes \qqttgag
and \ggttgag in perturbation theory, according to the
Kinoshita-Lee-Nauenberg (KLN) theorem\cite{KLN}, we should
consider these subprocesses too. The LO Feynman diagrams for the
partonic processes \qgttgaq$(q=u,d,s)$ can be obtained by
exchanging initial (anti)quark and final gluon in corresponding
diagrams in Fig.\ref{fig2a}.

\par
In order to avoid the additional IR singularity at the LO for the
partonic processes \qgttgaq$(q=u,d,s)$ due to the radiated photon
from a massless light-(anti)quark, we take a photon transverse
momentum cut and an angle cut between the jet and photon, e.g.
$p_T^{(\gamma)}>20~GeV$ and
$\theta_{\gamma,jet}>\theta_{\gamma,jet}^{cut}=3^\circ$[in the
proton-(anti)proton center-of-mass system.]$\footnote{From the
experimental point of view, we should apply angle cut
$\theta_{\gamma,jet}^{cut}$ not only to the $pp \to q(\bar q)g \to
t\bar t\gamma q(\bar q) + X$ processes, but also the $pp \to q\bar
q,gg \to t\bar t\gamma g + X$ processes. The subscript 'jet' in
$\theta_{\gamma,jet}$ represents the light-(anti)quark jet or gluon
jet for the real light-(anti)quark emission processes or the real
gluon emission processes.}$. Then these partonic processes contain
only the initial state collinear singularities induced by strong
interaction. Splitting the phase-space into collinear and
noncollinear regions by introducing a cutoff $\delta_c$, we can
express the cross sections for the partonic processes $qg \to t\bar
t\gamma q$ and $\bar{q}g \to t\bar t\gamma \bar{q}$ as
\begin{equation}
\hat{\sigma}^R(q(\bar{q})g \to t\bar t\gamma q(\bar q)) =
\hat{\sigma}^R_{q(\bar{q})g} = \hat{\sigma}^{C}_{q(\bar{q})g} +
\hat{\sigma}^{\overline{C}}_{q(\bar{q})g}
\end{equation}
The cross sections in the noncollinear region,
$\hat{\sigma}^{\overline{C}}_{q(\bar{q})g}$, are finite and can be
evaluated in four dimensions by using Monte Carlo method. The
differential cross sections in the collinear region for the $pp
\to q(\bar q)g \to t\bar t\gamma q(\bar q) + X,(q=u,d,s)$
processes, $d\sigma^{C}_{q(\bar{q})}$, can be written as
\begin{eqnarray}
d\sigma^{C}_{q} &=&
    \left[ \frac{\alpha_s}{2 \pi}
           \frac{\Gamma \left( 1 - \epsilon \right)}{\Gamma \left( 1 - 2 \epsilon \right)}
           \left( \frac{4 \pi \mu_r^2}{\hat{s}}
           \right)^{\epsilon}
    \right]
    \left( -\frac{1}{\epsilon} \right) \delta_c^{-\epsilon}
    \int_0^1 dx_1 \int_0^1 dx_2  \nonumber \\
 && \Big\{
    \Big[
    \int_{x_1}^{1} \frac{dz}{z} \left( \frac{1-z}{z} \right)^{-\epsilon}
    P_{qg} \left( z,\epsilon \right)
    G_{g/P_1} \left( x_1/z, \mu_f \right) G_{q/P_2} \left( x_2, \mu_f \right)
    d \hat{\sigma}^0_{\bar{q}q} \nonumber \\
&&  +~
    \int_{x_2}^{1} \frac{dz}{z} \left( \frac{1-z}{z} \right)^{-\epsilon}
    P_{qg} \left( z,\epsilon \right)
    G_{q/P_1} \left( x_1, \mu_f \right) G_{g/P_2} \left( x_2/z, \mu_f \right)
    d \hat{\sigma}^0_{q\bar{q}}
    \Big]  \nonumber \\
 &&+ \Big[
    \int_{x_1}^{1} \frac{dz}{z} \left( \frac{1-z}{z} \right)^{-\epsilon}
    P_{gq} \left( z,\epsilon \right)
    G_{q/P_1} \left( x_1/z, \mu_f \right) G_{g/P_2} \left( x_2, \mu_f \right) \nonumber \\
 &&+~ \int_{x_2}^{1} \frac{dz}{z} \left( \frac{1-z}{z} \right)^{-\epsilon}
    P_{gq} \left( z,\epsilon \right)
    G_{g/P_1} \left( x_1, \mu_f \right) G_{q/P_2} \left( x_2/z, \mu_f \right)
    \Big]
    d \hat{\sigma}^0_{gg}
    \Big\},
\end{eqnarray}
and
\begin{eqnarray}
 d\sigma^{C}_{\bar{q}} = d\sigma^{C}_{q}\left( q \leftrightarrow \bar{q}
 \right).
\end{eqnarray}

where the splitting functions $P_{qg(gq)}(z,\epsilon)$ can be
written explicitly as\cite{Altarelli}
\begin{eqnarray}
P_{qg,gq}(z,\epsilon)=P_{qg,gq}(z)+ \epsilon P'_{qg,gq}(z), ~~~
P_{qg}(z)=\frac{1}{2}[z^2+(1-z)^2],~~~ P'_{qg}(z) = -z(1-z), \nb
\end{eqnarray}
\begin{eqnarray}
P_{gq}(z)=C_F\frac{1+(1-z)^2}{z},~~~ P'_{gq}(z) = -C_Fz.
\end{eqnarray}

\vskip 5mm {\bf C. Virtual corrections  } \vskip 5mm

\begin{figure*}
\begin{center}
\includegraphics*[130pt,610pt][530pt,720pt]{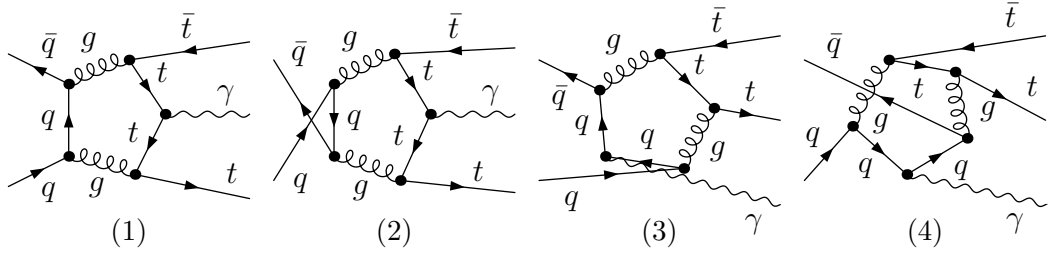}
\vspace*{-0.3cm} \centering \caption{\label{fig2d} The pentagon
Feynman diagrams for the partonic process \qqttga($q\bar q=u\bar
u,d\bar d$). }
\end{center}
\end{figure*}

\begin{figure*}
\begin{center}
\includegraphics*[130pt,410pt][530pt,720pt]{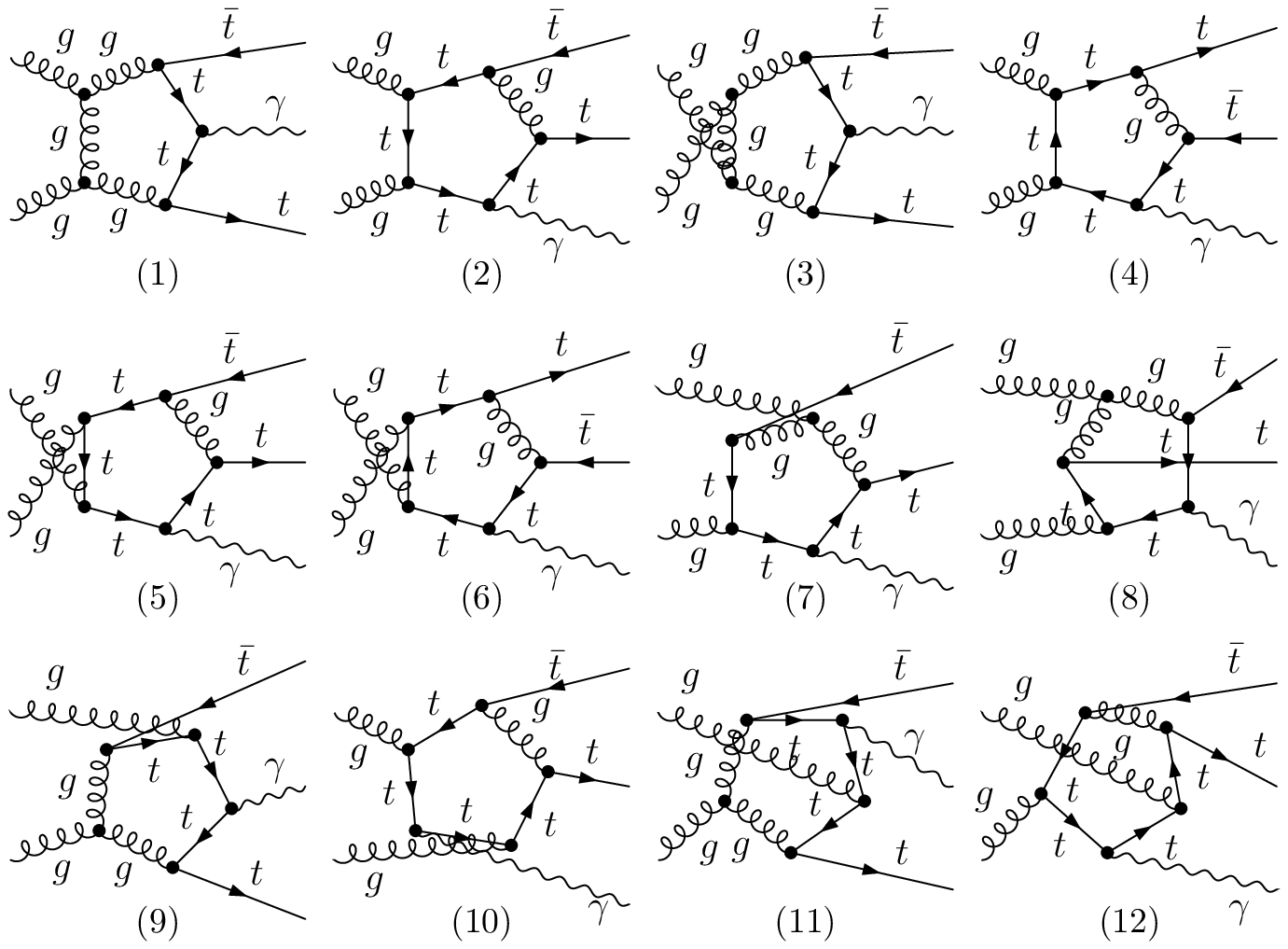}
\vspace*{-0.3cm} \centering \caption{\label{fig2e} The pentagon
Feynman diagrams for the partonic process \ggttga. }
\end{center}
\end{figure*}

There are 118 diagrams for the partonic process \qqttga in the SM
at NLO. It involves self-energy(40), vertex(32), box(14),
pentagon(4) and counterterm(28) graphs. For the partonic process
\ggttga there are 306 diagrams at NLO in the SM, including
self-energy(32), vertex(156), box(66), pentagon(12) and
counterterm(40) graphs. Among all these graphs the pentagon
diagrams are the most complicated ones. We depict them in
Fig.\ref{fig2d}(for partonic process \qqttga) and
Fig.\ref{fig2e}(for partonic process \ggttga). In the NLO
calculations, we use the dimensional regularization method and
adopt the modified minimal subtraction ($\overline{\rm MS}$)
scheme to renormalize the strong coupling constant and the
relevant masses and fields except for top-quark and gluon, where
their masses and wave functions are renormalized by applying the
on-shell scheme. The total NLO QCD amplitudes of partonic
processes \qqttga and \ggttga are UV finite after performing the
renormalization procedure. Nevertheless, they still contain
soft/collinear IR singularities.
%%figure%%
%%%%figure%%%%

\par
The virtual corrections to the subprocesses \qqttga and \ggttga can
be expressed as
\begin{eqnarray}
 && d\hat{\sigma}^V_{q\bar q}
    =
    \frac{1}{4}\frac{1}{9}\frac{(2 \pi)^4}{2\hat{s}}
    \sum_{spin}^{color}2 Re
    \left[
    {\cal M}_{LO}^{q\bar q}{\cal M}^V_{q\bar q}
    \right] d \Omega_{3}^{q\bar q} \nonumber \\
 && d\hat{\sigma}^V_{gg}
    =
    \frac{1}{4}\frac{1}{64}\frac{(2 \pi)^4}{2\hat{s}}
    \sum_{spin}^{color}2 Re
    \left[{\cal M}_{LO}^{gg}{\cal M}^V_{gg}
    \right]d \Omega_{3}^{gg}
\end{eqnarray}
where ${\cal M}_{LO}^{q\bar q}$ and ${\cal M}_{LO}^{gg}$ are the LO
Feynman matrices of the partonic processes \qqttga and \ggttga, and
${\cal M}^{V}_{q\bar q}$ and ${\cal M}^{V}_{gg}$ are the NLO
matrices for the $q-\bar q$ and $g-g$ annihilation processes,
separately.

\par
The virtual correction parts of the cross sections containing
soft/collinear IR singularities. As we can see later that the
soft/collinear IR singularities are canceled exactly after combining
the virtual corrections to the pertonic processes $pp(p\bar p) \to
q\bar q(gg) \to t\bar t\gamma+X$ with the real gluon emission
corrections and the gluon part of the PDF counterterms $\delta
G_{q(g)/P}^{(g)}~(P=p,\bar{p}~;~ q = u, \bar{u}, d, \bar{d}, s,
\bar{s})$.

\vskip 5mm {\bf D. NLO QCD corrected cross section for \ppttga
process } \vskip 5mm
\par
As shown in Eqs.(\ref{PDFcounterterm1}), the PDF counterterms
contain collinear IR singularities. By combining the contributions
of the PDF counterterms with the hard collinear contributions of
the \qqttgag, \ggttgag, \qgttgaq subprocesses, we get the
expression for the remaining collinear contributions to the
process \ppttga in ${\cal O}(\alpha_s)$ order as,
\begin{eqnarray} \label{collinear
cross section1}
  d \sigma^{coll}
  &=&
  \left[ \frac{\alpha_s}{2 \pi}
           \frac{\Gamma \left( 1 - \epsilon \right)}{\Gamma \left( 1 - 2 \epsilon \right)}
           \left( \frac{4 \pi \mu_r^2}{\hat{s}}
           \right)^{\epsilon}
  \right]
  \sum_{ij = u\bar{u}, d\bar{d}}^{s\bar{s},gg} \frac{1}{1+\delta_{ij}}
  \int_0^1 dx_1 \int_0^1 dx_2 \nonumber \\
 && \times \Big\{
    \Big[
    \tilde{G}_{i/P_1} \left( x_1,\mu_f \right) G_{j/P_2} \left( x_2,\mu_f \right)
    +G_{i/P_1} \left( x_1,\mu_f \right) \tilde{G}_{j/P_2} \left( x_2,\mu_f \right) \nonumber \\
 && +\sum_{\alpha=i,j}
     \Big(
          \frac{A_1^{sc}(\alpha \rightarrow \alpha g)}{\epsilon}
          +A_0^{sc}(\alpha \to \alpha g)
     \Big)
     \Big]
     d\hat{\sigma}^0_{ij} \nonumber \\
  && +\left( i \leftrightarrow j \right)
     \Big\}
\end{eqnarray}

where
\begin{eqnarray}
A_1^{sc}(q \to qg)= C_F(2 \ln \delta_s+3/2), ~~~~A_1^{sc}(g\to
gg)=2N\ln\delta_s+(11N-2n_{lf})/6,
\end{eqnarray}
\begin{eqnarray}
A_0^{sc} = A_1^{sc} \ln(\frac{\hat{s}}{\mu_f^2}),~~~~
\tilde{G}_{\alpha/P}(x, \mu_f) = \sum_{\alpha^{\prime}}
\int_{x}^{1-\delta_s \delta_{\alpha \alpha^{\prime}}} \frac{dy}{y}
G_{\alpha^{\prime}/P}(x/y, \mu_f) \tilde{P}_{\alpha
\alpha^{\prime}}(y),
\end{eqnarray}
and
\begin{eqnarray}
\tilde{P}_{\alpha \alpha^{\prime}}(y) =
  P_{\alpha \alpha^{\prime}}(y)
  \ln \left(
     \delta_c \frac{1-y}{y} \frac{\hat{s}}{\mu_f^2}
     \right)
  -P^{\prime}_{\alpha \alpha^{\prime}}(y),
\end{eqnarray}
where $N=3$ and $n_{lf}=5$, respectively. The explicit expressions
for $P_{\alpha \alpha^{\prime}}$ and $P^{\prime}_{\alpha
\alpha^{\prime}}$ can be found in Ref.\cite{Harris}. By adding the
virtual correction, the soft real gluon emission corrections and the
remaining collinear contributions shown in Eq.(\ref{collinear cross
section1}), the soft and collinear IR divergences are vanished. The
final result for the total QCD correction($\Delta\sigma^{QCD}$)
consists of a three-body term and a four-body term, i.e.,
$\Delta\sigma^{QCD}=\Delta\sigma^{(3)}+\Delta\sigma^{(4)}$. The
three-body term can be expressed as
\begin{eqnarray} \label{3body}
 && \Delta \sigma^{(3)}= \int d \sigma^{coll}+ \sum_{ij =
 u \bar{u}, d \bar{d}}^{s\bar{s}, c\bar{c}, gg} \frac{1}{1+\delta_{ij}}  \nonumber \\
  &&  \times  \int_0^1 dx_1 \int_0^1 dx_2
  \int  \Big[  \left(
  d \hat{\sigma}_{ij}^V + d \hat{\sigma}_{g,ij}^S
  \right)
  G_{i/P_1}\left( x_1,\mu_f \right) G_{j/P_2} \left( x_2,\mu_f \right)
  + \left( i \leftrightarrow j \right)
  \Big],
\end{eqnarray}

and the four-body term has the form as
\begin{eqnarray}\label{4body}
  \Delta \sigma^{(4)}
  &=&
  \sum_{ij = u\bar{u}, d\bar{d}}^{s\bar{s}, gg} \frac{1}{1+\delta_{ij}}
  \int_0^1 dx_1 \int_0^1 dx_2
    \Big[
    G_{i/P_1} \left( x_1,\mu_f \right) G_{j/P_2} \left( x_2,\mu_f \right)
    \hat{\sigma}_{g,ij}^{\overline{\rm HC}}
    +\left( i \leftrightarrow j \right)
    \Big] \nonumber \\
 && +
  \int_0^1 dx_1 \int_0^1 dx_2
    \Big[
    G_{c/P_1} \left( x_1,\mu_f \right) G_{\bar{c}/P_2} \left( x_2,\mu_f \right)
    \hat{\sigma}_{g,c\bar{c}}^{H}
    +\left( c \leftrightarrow \bar{c} \right)
    \Big] \nonumber \\
 && +
   \sum_{q=u,d,s}^{\bar{u}, \bar{d}, \bar{s}}
   \int_0^1 dx_1 \int_0^1 dx_2
   \Big[
   G_{q/P_1} \left( x_1,\mu_f \right) G_{g/P_2} \left( x_2,\mu_f \right)
   \hat{\sigma}_{qg}^{\overline{\rm C}}
   +\left( q \leftrightarrow g \right)
   \Big],
\end{eqnarray}
where $\hat{\sigma}^H_{g,c\bar{c}}(\hat{s} = x_1 x_2 s)$ and
$\hat{\sigma}^{\overline{\rm HC}}_{g,ij}(\hat{s}=x_1 x_2 s)$ are the
cross sections for the partonic processes $c\bar{c} \to t \bar{t}
\gamma g$ and $ij \to t\bar t\gamma g~(ij=u\bar u,d\bar d,s\bar
s,gg)$ in the hard and hard noncollinear phase-space regions at the
colliding energy $\hat{s}=x_1 x_2 s$ respectively, and
$\hat{\sigma}_{qg}^{\overline{C}}(\hat{s})~(q=u, \bar{u}, d,
\bar{d}, s, \bar{s})$ represent the cross sections in the
noncollinear phase-space region for the partonic processes $u g \to
t\bar t\gamma u$, $d g \to t\bar t\gamma d$, $\bar u g \to t\bar
t\gamma \bar u$, $\bar d g \to t\bar t\gamma \bar d$, $s g \to t
\bar{t} \gamma s$ and $\bar{s} g \to t \bar{t} \gamma \bar{s}$,
respectively.

\par
Finally, the QCD corrected total cross section for the \ppttga
process is
\begin{eqnarray}\label{sigmaQCD}
\sigma^{QCD}=\sigma^{0}+\Delta\sigma^{QCD}=\sigma^{0}+\Delta\sigma^{(3)}+\Delta\sigma^{(4)}.
\end{eqnarray}

\par
In adopting Eqs.(\ref{3body}), (\ref{4body}), and (\ref{sigmaQCD})
for the numerical calculation, we use the CTEQ6M\cite{pdfs} PDFs.

\par
\section{Numerical results and discussion}
\par
In this section we describe and discuss the numerical results for
the LO and NLO QCD corrected physical observables for the
processes \ppttga. We take one-loop and two-loop running
$\alpha_{s}$ in the LO and NLO calculations,
respectively\cite{hepdata}. The number of active flavors is
$N_f=5$, and the QCD parameters are $\Lambda_5^{LO}=165~MeV$ and
$\Lambda_5^{\overline{MS}}=226~MeV$ for the LO and NLO
calculations, respectively. We set the factorization scale and the
renormalization scale being equal(i.e., $\mu=\mu_f=\mu_r$) and
take $\mu=\mu_0=m_t$ by default unless otherwise stated.
Throughout this paper, we take $m_t=171.2~GeV$ and
$\alpha(m_Z)^{-1}=127.918$\cite{hepdata}, and set quark masses
$m_u=m_d=m_s=0$ and $m_c=1.3~GeV$ which are the same as the input
parameters used in CTEQ PDFs\cite{pdfs}. The colliding energies in
the proton-(anti)proton center-of-mass system are taken as $\sqrt
s=14~TeV$ for the LHC and $\sqrt s=1.96~TeV$ for the Tevatron Run
II.

\par
To distinguish the photon from the jets requires the angle between
the outgoing photon and jet constrained in the range of
$\theta_{\gamma,jet}>\theta_{\gamma,jet}^{cut}$(in the
center-of-mass system of proton-(anti)proton). In our calculation we
assume the produced (anti)top-quarks are always tagged and not
effected by the phase-space cut. During our numerical calculation,
we applied a number of checks to our calculations.

\par
1. We have compared the numerical results of the LO cross section
for the process $pp \to d\bar d \to t\bar t\gamma+X$ with
$\sqrt{s}=14~TeV$ by using FeynArts3.4/FormCalc5.4
\cite{fey,formloop} packages and CompHEP-4.4p3
program\cite{CompHEP}, and applying the Feynman and unitary gauges,
separately. In the partonic luminosity integrations we adopt the
CTEQ6L1 PDFs. The results are $125.1(1)~fb$ (CompHEP, Feynman
gauge), $125.0(1)~fb$ (CompHEP, unitary gauge), $125.1(1)~fb$
(FeynArts, Feynman gauge) and $125.1(1)~fb$ (FeynArts, unitary
gauge). It demonstrates all these results are in good agreement.

\par
2. We checked the UV cancelation both analytically and numerically.
The IR finiteness is verified numerically after combining all the
contributions from the virtual corrections, renormalization
constants, real gluon/light-(anti)quark emission partonic processes
and the collinear counterterms of PDFs at the NLO.

\par
3. We used both the LoopTools2.2 package\cite{formloop} and our
in-house library to check the correctness of the deductions of the
tensor integrals and IR-finite numerical calculation of the scalar
integral functions at a few phase-space points. Both packages are
developed based on the same expressions given in
Refs.\cite{OneTwoThree,Four,Five}, but in different codes. The
numerical results are coincident within the calculation errors.

\par
\begin{figure}[htbp]
\includegraphics[scale=1.2]{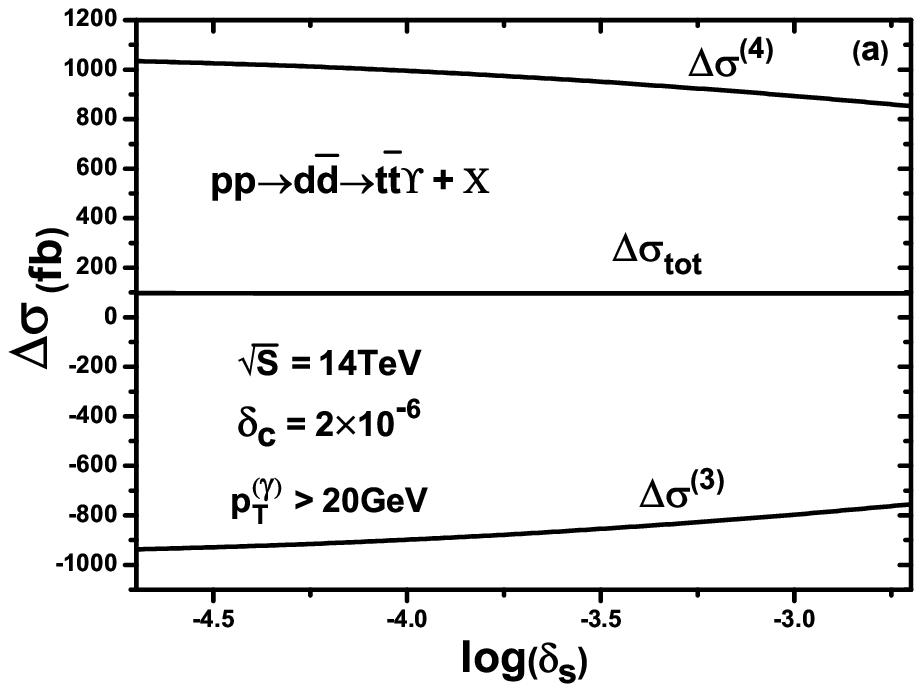}
\includegraphics[scale=1.2]{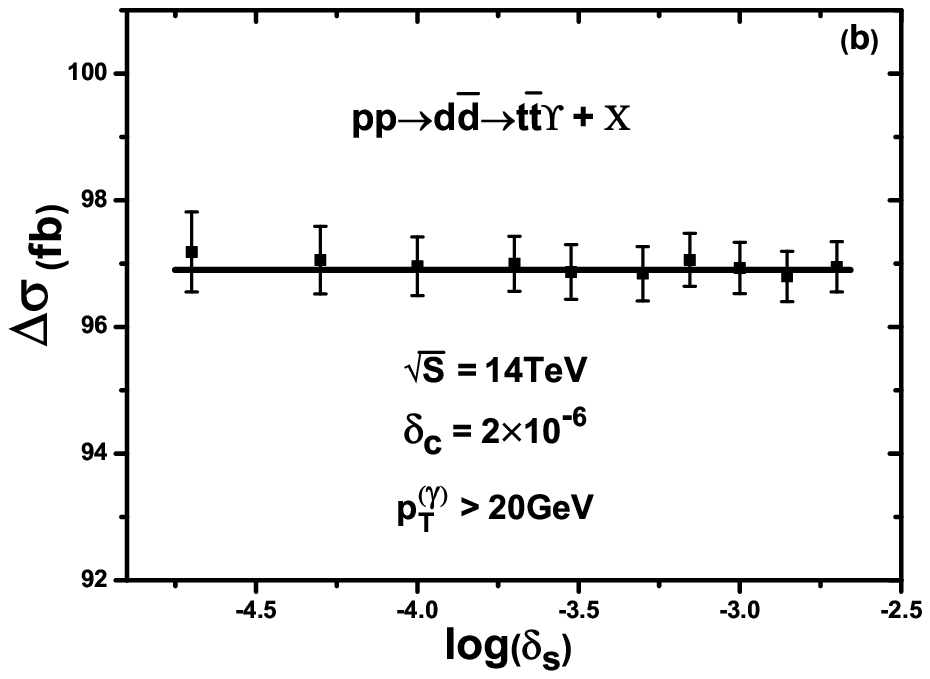}
\hspace{0in}%
\caption{\label{fig3a} (a) The dependence of the NLO QCD correction
parts to the \ppddttga process on the soft cutoff $\delta_s$ at the
LHC. (b) The amplified curve for the total NLO QCD correction
$\Delta\sigma^{QCD}$ to the process \ppddttga in Fig.\ref{fig3a}(a),
where it includes the calculation errors. }
\end{figure}
\begin{figure}[htbp]
\includegraphics[scale=1.2]{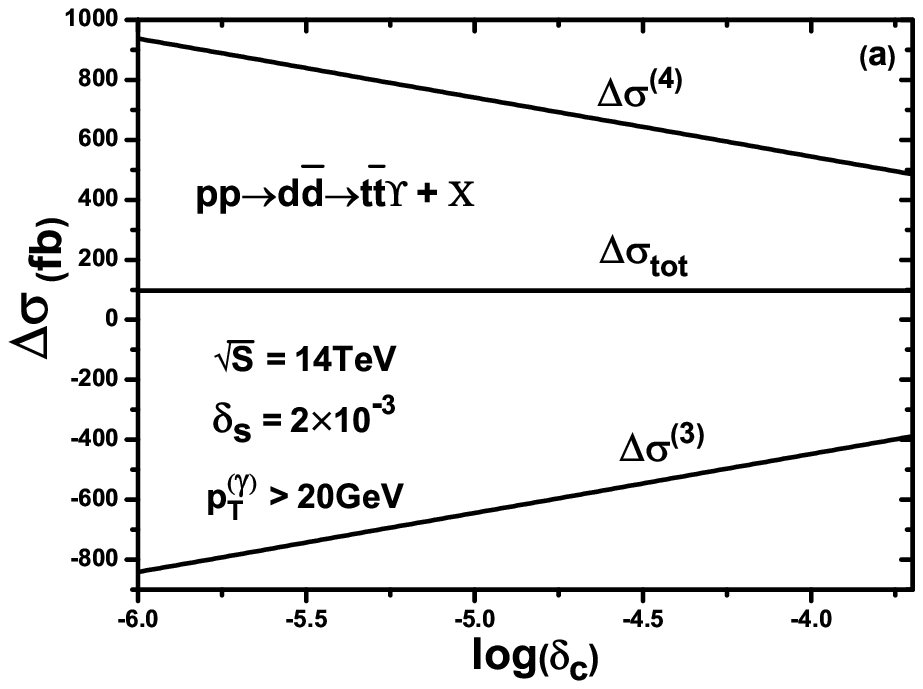}
\includegraphics[scale=1.2]{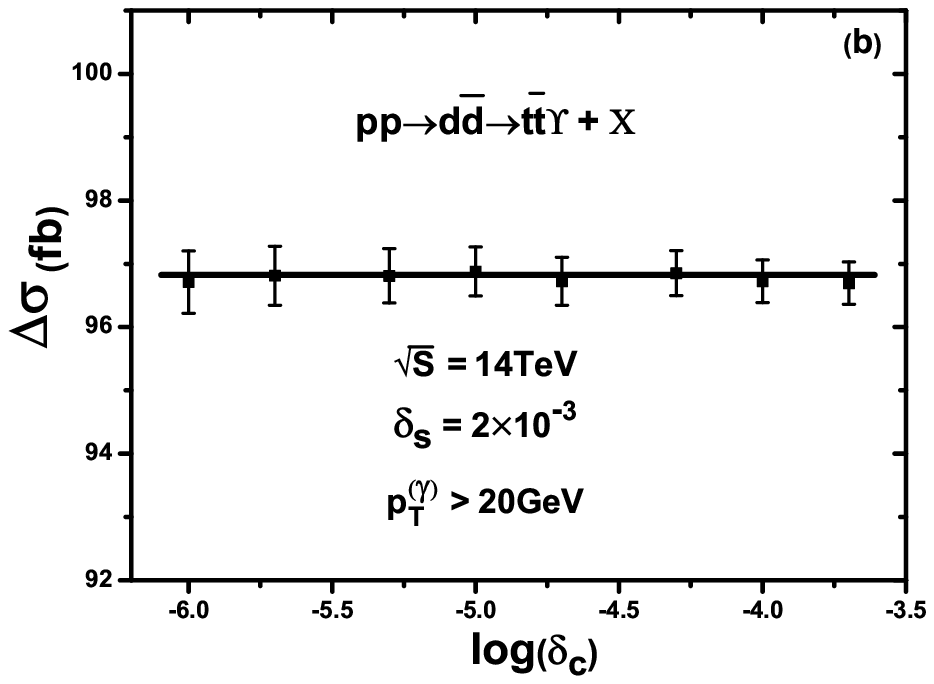}
\caption{\label{fig4a} (a) The dependence of the NLO QCD correction
parts to the \ppddttga process on the collinear cutoff $\delta_c$ at
the LHC. (b) The amplified curve for the total QCD correction
$\Delta\sigma^{QCD}$ to the process \ppddttga in Fig.\ref{fig4a}(a),
where it includes the calculation errors.}
\end{figure}

4. The independence of the NLO correction on the soft cutoff
$\delta_s$ and collinear cutoff $\delta_c$, were proofed. Figures
\ref{fig3a}(a) and \ref{fig3a}(b) demonstrate that the total NLO QCD
correction to the \ppddttga process at the LHC does not depend on
the arbitrarily chosen value of the cutoff $\delta_s$ within the
calculation errors, where we take the photon transverse momentum cut
$p_T^{(\gamma)}>20~GeV$, $\theta_{\gamma,jet}>3^{\circ}$[in the
proton-(anti)proton center-of-mass system] and $\delta_c = 2 \times
10^{-6}$ in adopting the TCPSS method. In Fig.\ref{fig3a}(a), the
three-body correction[$\Delta\sigma^{(3)}$, see Eq.(\ref{3body})],
four-body correction[$\Delta\sigma^{(4)}$, see Eq.(\ref{4body})],
and the total QCD correction ($\Delta\sigma^{QCD}$) for the
\ppddttga process, are depicted as the functions of the soft cutoff
$\delta_s$ with $\delta_s$ running from $2\times 10^{-5}$ to
$2\times 10^{-3}$. The amplified curve for the total QCD correction,
$\Delta\sigma^{QCD}$, is presented in Fig.\ref{fig3a}(b) together
with calculation errors. Figures \ref{fig4a}(a) and \ref{fig4a}(b)
show the independence of the total NLO QCD correction to the
\ppddttga process on the cutoff $\delta_c$ where we take $\delta_s=2
\times 10^{-3}$, $\theta_{\gamma,jet}>3^{\circ}$ and
$p_T^{(\gamma)}>20~GeV$. In Fig.\ref{fig4a}(b) the amplified curve
for $\Delta\sigma^{QCD}$ for the \ppddttga process is depicted. The
verification that the total QCD correction $\Delta\sigma^{QCD}$ for
the \ppddttga process is independent of these two cutoffs, not only
demonstrates the cancellation of soft/collinear IR divergencies in
the total NLO QCD corrections to the processes \ppddttga, but also
provide an indirect check for the correctness of our calculations.
In further numerical calculations, we fix $p_T^{(\gamma)}>20~GeV$,
$\theta_{\gamma,jet}^{cut}=3^\circ$, $\delta_s=10^{-3}$ and
$\delta_c=\delta_s/50$, if there is no other statement.
%%%%figure%%%%
%%%%figure%%%%
%%%%figure%%%%
%%%%figure%%%%

\par
\begin{figure}
\centering
\includegraphics[scale=1.0]{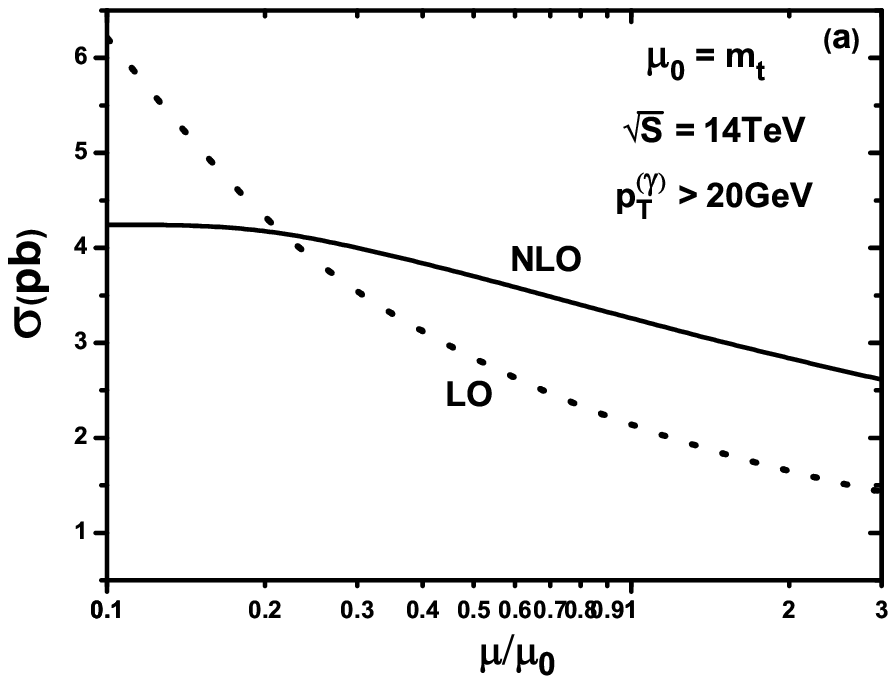}
\includegraphics[scale=1.0]{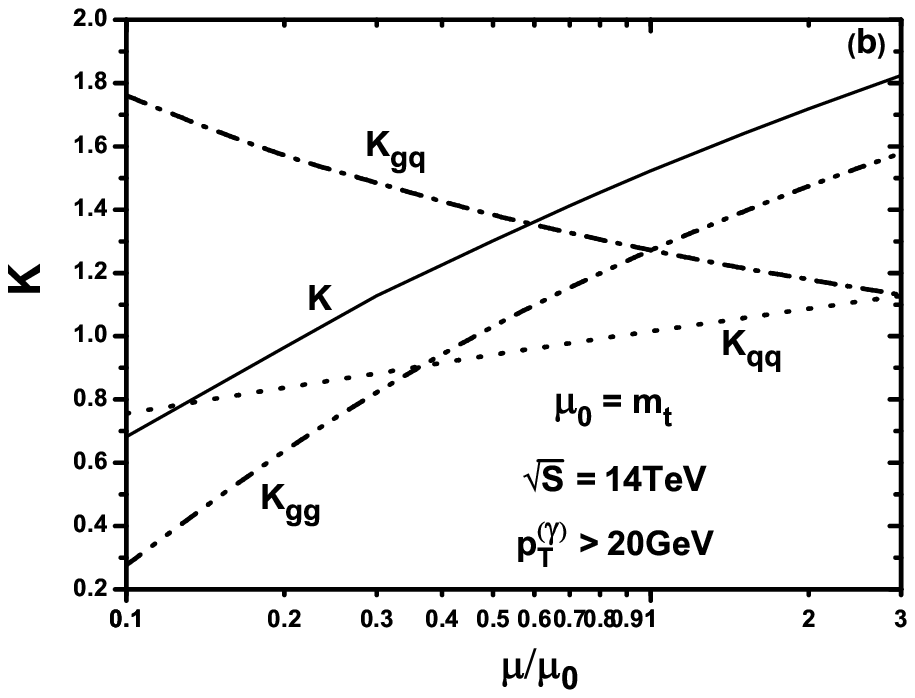}
\caption{\label{fig5a} (a) The dependence of the LO and NLO cross
sections on the factorization/renormalization scale at the LHC.
(b)The total NLO QCD K-factor for the process[$ K\equiv
\sigma^{QCD}/\sigma_{LO}$], the NLO QCD K-factor from the $p\bar p
\to q\bar q \to t\bar t \gamma+X(q=u,d,s,c)$ processes [$K_{q\bar
q}\equiv 1+\frac{\sum_{q=u,d,}^{s,c}(\Delta \sigma_{q\bar
q}^{QCD})}{\sigma_{LO}}$], the $p\bar p \to gg \to t\bar t \gamma+X$
process [$K_{gg}\equiv 1+ \Delta \sigma_{gg}^{QCD}/\sigma_{LO}$] and
the $p\bar p \to gq(\bar q) \to t\bar t \gamma+X,~(q=u,d,s)$
processes [$K_{gq}\equiv 1+\frac{\sum_{q=u,d,}^{s}(\Delta
\sigma_{gq}^{QCD}+\Delta \sigma_{g\bar q}^{QCD})}{\sigma_{LO}}$]
versus the energy scale at the LHC.  }
\end{figure}
\begin{figure}
\centering
\includegraphics[scale=1.0]{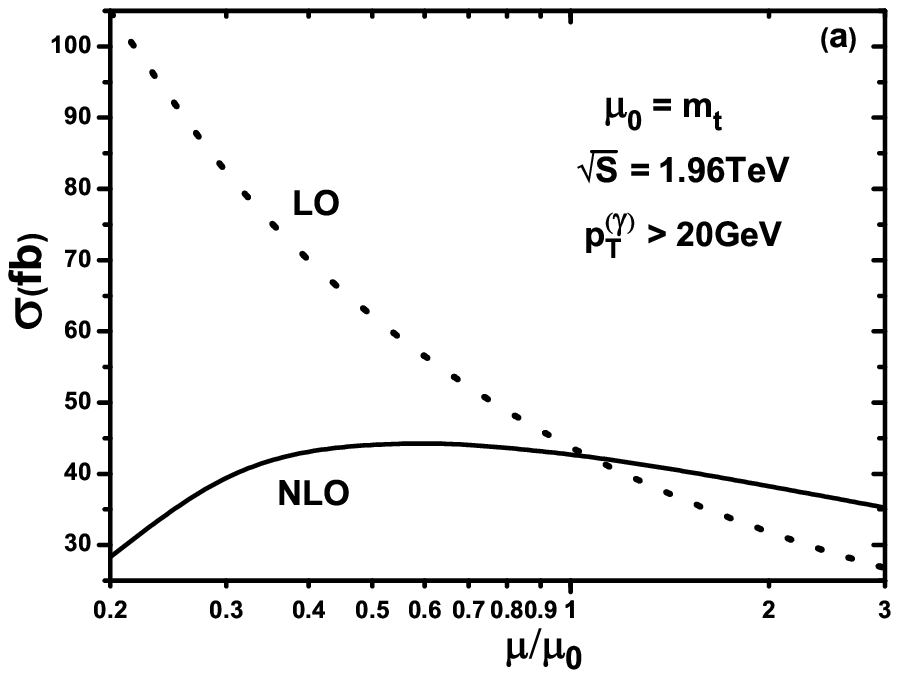}
\includegraphics[scale=1.0]{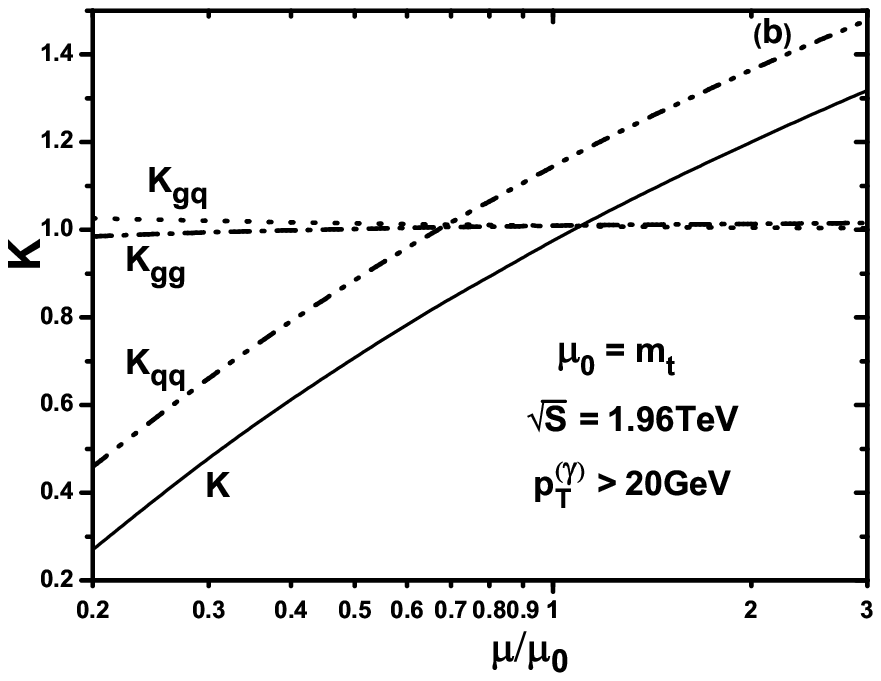}
\caption{\label{fig6a} (a) The dependence of the LO and NLO cross
sections on the factorization/renormalization scale at the Tevatron.
(b) The total NLO QCD K-factor for the process($K\equiv \Delta
\sigma^{QCD}/\sigma_{LO}$), the NLO QCD K-factor from the $p\bar p
\to q\bar q \to t\bar t \gamma+X(q=u,d,s,c)$ processes [$K_{q\bar
q}\equiv 1+\frac{\sum_{q=u,d,}^{s,c}(\Delta \sigma_{q\bar
q}^{QCD})}{\sigma_{LO}}$], the $p\bar p \to gg \to t\bar t \gamma+X$
process [$K_{gg}\equiv 1+ \Delta \sigma_{gg}^{QCD}/\sigma_{LO}$] and
the $p\bar p \to gq(\bar q) \to t\bar t \gamma+X,~(q=u,d,s)$
processes [$K_{gq}\equiv 1+\frac{\sum_{q=u,d,}^{s}(\Delta
\sigma_{gq}^{QCD}+\Delta \sigma_{g\bar q}^{QCD})}{\sigma_{LO}}$]
versus the energy scale at the Tevatron. }
\end{figure}

In Fig.\ref{fig5a}(a) and Figs.\ref{fig6a}(a) we present the
dependence of the integrated LO and the NLO QCD corrected cross
sections on the renormalization/factorization scale($\mu$) at the
LHC and Tevatron RUN II, separately. There we assume $\mu =
\mu_r=\mu_f$ and define $\mu_0 \equiv m_t$. We can see that although
the curves for the LO and NLO cross sections have visible variations
when the energy scale $\mu$ runs from $0.1 m_t$ to $3 m_t$, the
curves for NLO become more stable in comparison with the
corresponding curves for LO. It demonstrates that the NLO QCD
corrections reduce obviously the dependence of the cross section on
the introduced parameter $\mu$ in the plotted $\mu/\mu_0$ value
range. The corresponding total K-factor[$K \equiv
\sigma^{QCD}/\sigma_{LO}$], the NLO QCD K-factor from the \ppqqttga
processes[$K_{q\bar q}\equiv 1+\frac{\sum_{q=u,d,}^{s,c}(\Delta
\sigma_{q\bar q}^{QCD})}{\sigma_{LO}}$], the K-factor from the
\ppggttga process [$K_{gg}\equiv
1+\frac{\Delta\sigma_{gg}^{QCD}}{\sigma_{LO}}$] and the K-factor
from the \ppgqttga processes [$K_{gq}\equiv
1+\frac{\sum_{q=u,d,}^{s}(\Delta \sigma_{gq}^{QCD}+\Delta
\sigma_{g\bar q}^{QCD})}{\sigma_{LO}}$] are plotted in
Fig.\ref{fig5a}(b) for the LHC and Fig.\ref{fig6a}(b) for the
Tevatron. Fig.\ref{fig5a}(b) shows that at the LHC the integrated
NLO QCD corrections always enhance the LO cross sections except in
the range of $0.1<\mu/\mu_0<0.21$, and the NLO QCD corrections from
the $pp \to gg \to t\bar t\gamma+X$ and $pp \to q\bar q \to t\bar
t\gamma+X(q=u,d,s,c)$ processes counteract the other contributions
in the region of $0.1<\mu/\mu_0<0.46$. Fig.\ref{fig5a}(b) shows when
we take $\mu/\mu_0=1$, the NLO QCD correction to the $pp \to gg \to
t\bar t\gamma+X$ process at the LHC is much larger than the
correction to $pp \to q\bar q \to t\bar t\gamma+X$ processes, while
Figure \ref{fig6a}(b) demonstrates the NLO QCD correction to the
$p\bar p \to q\bar q \to t\bar t\gamma+X$ processes at the Tevatron
is larger than that to $p\bar p \to gg \to t\bar t\gamma+X$ in the
vicinity of $\mu/\mu_0=1$. In the further calculations, we fix
$\mu=\mu_0=m_t$.

%%figure%%
%%%%figure%%%%
%%figure%%
%%%%figure%%%%

\par
In Table \ref{tab1} we list the numerical results related to the
data in Figs.\ref{fig5a}(a,b) for the LHC and Figs.\ref{fig6a}(a,b)
for the Tevatron at the position of $\mu=\mu_0$. They are the
results for the integrated LO and NLO QCD corrected cross sections,
the total K-factor[$K\equiv \frac{\sigma^{QCD}}{\sigma_{LO}}$] of
the process \ppttga, the K-factor contributed by all the NLO QCD
corrections to the \ppqqttga$(q=u,d,s,c)$ processes [$K_{q\bar
q}\equiv 1+\frac{\sum_{q=u,d,}^{s,c}(\Delta \sigma_{q\bar
q}^{QCD})}{\sigma_{LO}}$], the K-factor contributed by the
integrated NLO QCD correction to the \ppggttga process
[$K_{gg}\equiv 1+\frac{\Delta \sigma_{gg}^{QCD}}{\sigma_{LO}}$] and
the K-factor contributed by the corrections to the \ppgqttga
processes [$K_{gq}\equiv 1+\frac{\sum_{q=u,d,}^{s}(\Delta
\sigma_{gq}^{QCD}+\Delta \sigma_{g\bar q}^{QCD})}{\sigma_{LO}}$] at
the LHC and Tevatron.

\begin{table}
\begin{center}
\begin{tabular}{|c|c|c|c|c|c|c|}
\hline Collider & $\sigma_{LO}$ & $\sigma^{QCD}$ &
$K_{q\bar q}$ & $K_{gg}$  & $K_{gq}$ & $K $  \\
\hline   LHC    & 2.141(2) (pb) & 3.256(6) (pb)  & 1.014  & 1.272  & 1.272 & 1.521    \\
\hline Tevatron & 43.79(4) (fb) & 42.80(6) (fb)  & 1.148  & 1.009  & 1.009 & 0.977    \\
\hline
\end{tabular}
\end{center}
\begin{center}
\begin{minipage}{15cm}
\caption{\label{tab1} The LO and NLO cross sections, the total
K-factor($K\equiv\frac{\sigma^{QCD}}{\sigma_{LO}}$), the K-factor
contributed by the NLO QCD correction to the \ppqqttga($q=u,d,s,c$)
processes[$K_{q\bar q}\equiv 1+\frac{\sum_{q=u,d,}^{s,c}(\Delta
\sigma_{q\bar q}^{QCD})}{\sigma_{LO}}$], the K-factor contributed by
the NLO QCD correction to the \ppggttga process [$K_{gg}\equiv
1+\frac{\Delta \sigma_{gg}^{QCD}}{\sigma_{LO}}$] and the K-factor
contributed by the corrections to the $pp(p\bar p) \to gq(g\bar q)
\to t\bar t \gamma q(\bar q)+X$($q=u,d,s$) processes [$K_{gq}\equiv
1+\frac{\sum_{q=u,d,}^{s}(\Delta \sigma_{gq}^{QCD}+\Delta
\sigma_{g\bar q}^{QCD})}{\sigma_{LO}}$] with $\mu=m_t$,
$p_T^{(\gamma)}>20~GeV$ and $\theta_{\gamma,jet}^{cut}=3^\circ$ at
the LHC and Tevatron. }
\end{minipage}
\end{center}
\end{table}

\par
\begin{figure}
\centering
\includegraphics[scale=1.0]{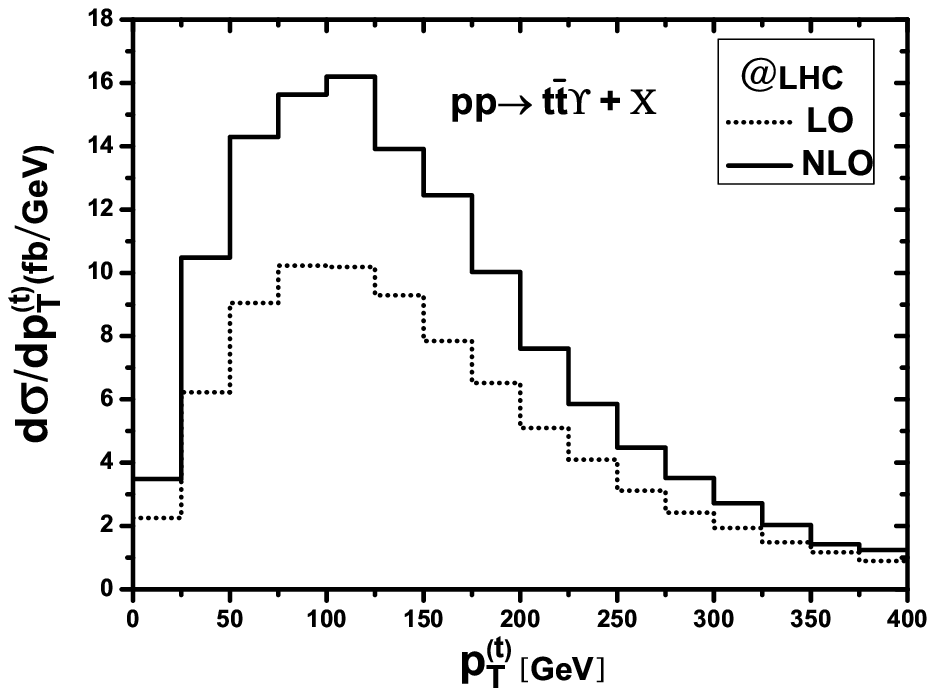}
\includegraphics[scale=1.0]{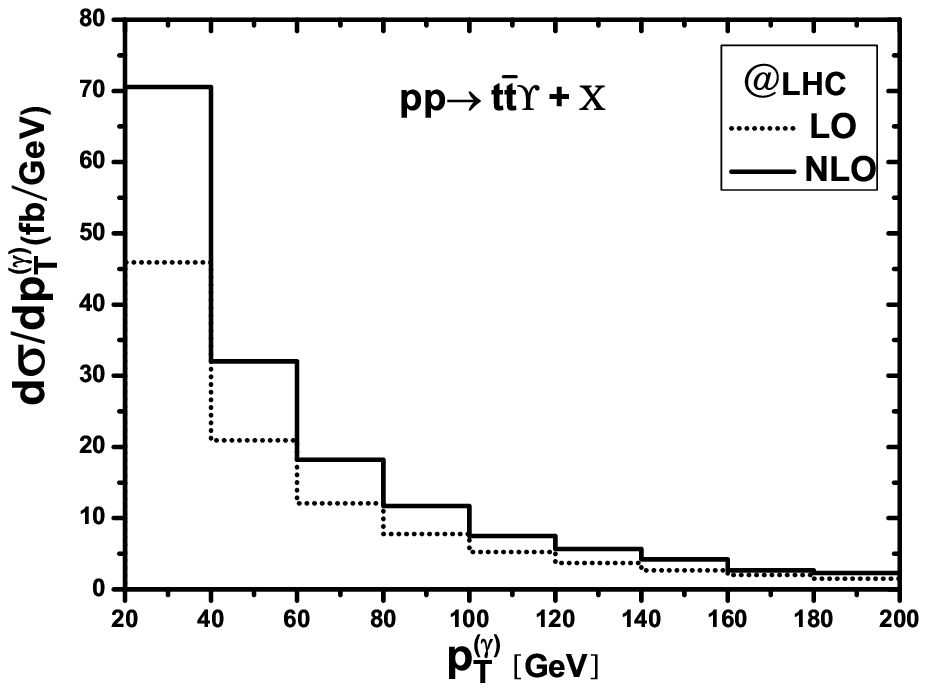}
\caption{\label{fig7a} The LO and NLO distributions of the
transverse momenta of the top quark and photon taking $\mu=m_t$,
$p_T^{(\gamma)}>20~GeV$ and $\theta_{\gamma,jet}^{cut}=3^\circ$ at
the LHC. (a) for the top quark, (b) for the photon. }
\end{figure}
\begin{figure}
\centering
\includegraphics[scale=1.0]{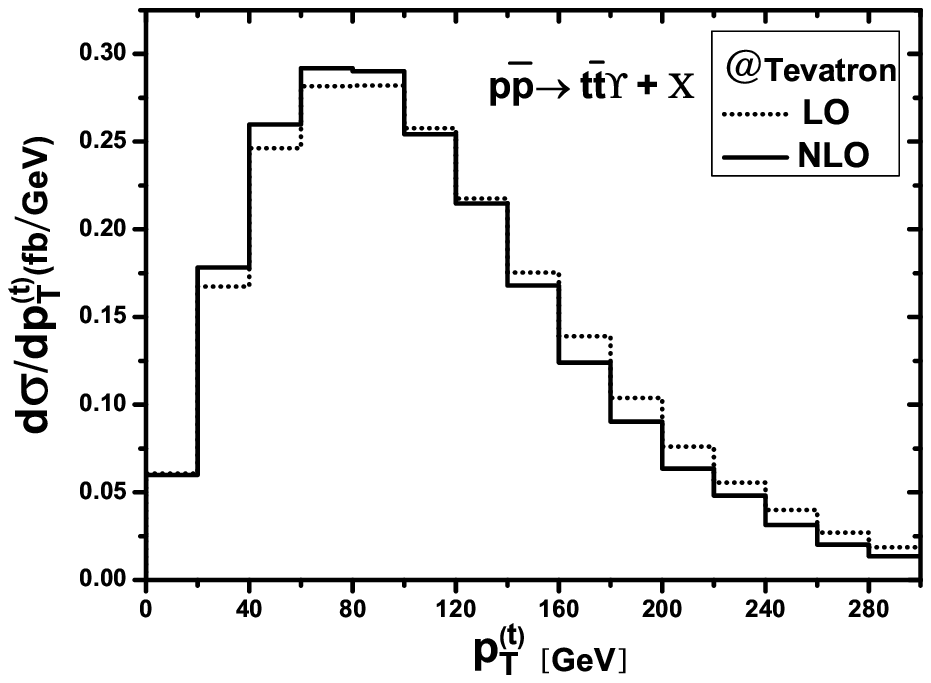}
\includegraphics[scale=1.0]{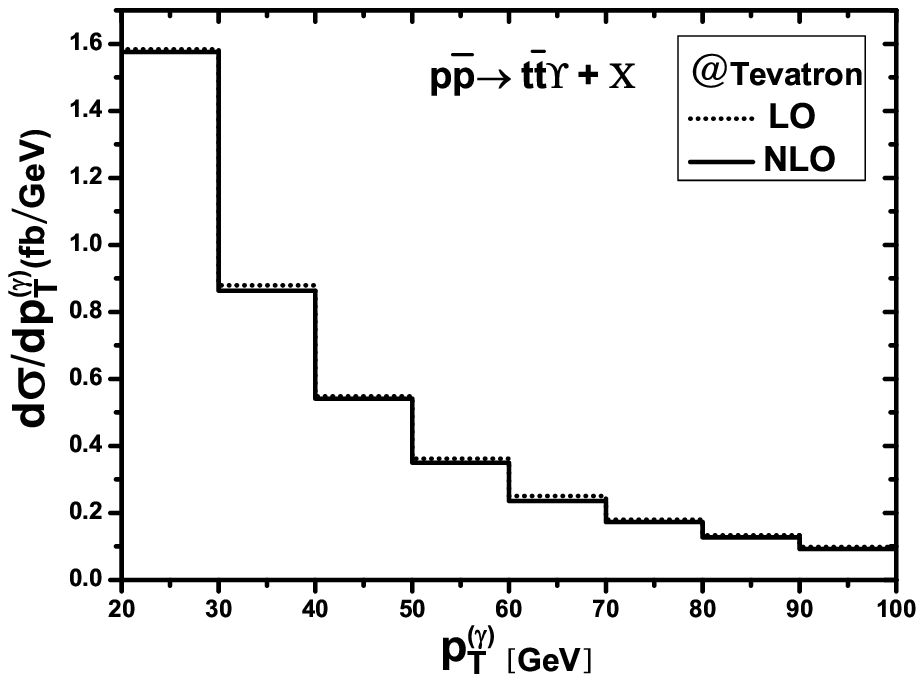}
\caption{\label{fig8a} The LO and NLO distributions of the
transverse momenta of the top quark and photon taking $\mu=m_t$,
$p_T^{(\gamma)}>20~GeV$ and $\theta_{\gamma,jet}^{cut}=3^\circ$ at
the Tevatron. (a) for the top quark, (b) for the photon. }
\end{figure}

The LO and NLO differential cross sections of the transverse momenta
for the top quark and photon at the LHC, are depicted in
Fig.\ref{fig7a}(a) and Fig.\ref{fig7a}(b), respectively. The
analogous plots at the Tevatron are depicted in Figs.\ref{fig8a}(a)
and (b). Figures \ref{fig7a}(a) and \ref{fig7a}(b) demonstrate that
the NLO QCD corrections enhance significantly the differential cross
sections of $p_T^{(t)}$ and $p_T^{(\gamma)}$ for the LHC, but the
corrections make only a small impact on the distributions of
$p_T^{(t)}$ and $p_T^{(\gamma)}$ for the Tevatron as shown in
Figs.\ref{fig8a}(a,b). In Figs.\ref{fig8a}(a,b) the NLO corrections
at the Tevatron suppress the LO distribution of $p_T^{(t)}$ a little
bit except in the range of $20~GeV<p_T^{(t)}<100~GeV$, while the
corrections slightly reduce the LO differential cross section of
$p_T^{(\gamma)}$ in the whole plotted $p_T^{(\gamma)}$ range. From
the distributions of $p_T^{(\gamma)}$ in both Fig.\ref{fig7a}(b) and
Fig.\ref{fig8a}(b), we can conclude that most of the photons in the
events of \ppttga are produced in low transverse momentum range at
the LHC and Tevatron.

\par
We adopt the definitions of the LO and NLO top-quark charge
asymmetries in Ref.\cite{Wein}, i.e.,
\begin{eqnarray}
\label{chargeAS}
A_{FB,LO}^t=\frac{\sigma_{LO}^-}{\sigma_{LO}^+},~~~~~
A_{FB,NLO}^t=\frac{\sigma_{LO}^-}{\sigma_{LO}^+}\left(1+\frac{\Delta
\sigma_{NLO}^-}{\sigma_{LO}^-}-\frac{\Delta
\sigma_{NLO}^+}{\sigma_{LO}^+}\right ),
\end{eqnarray}
In Eq.(\ref{chargeAS}) the notations $\sigma_{LO}^{\pm}$ have the
explicit definitions as
\begin{eqnarray}
\sigma_{LO}^{\pm}=\sigma_{LO}(y_t>0)\pm\sigma_{LO}(y_t<0),
\end{eqnarray}
where cross sections $\sigma_{LO}(y_t>0)$ and $\sigma_{LO}(y_t<0)$
get the contributions from the top-quarks in the forward and
backward hemispheres at LO, respectively,[The forward direction is
defined as the orientation for incoming proton($P_1$).],
$\Delta\sigma_{NLO}^{\pm}$ denote the NLO QCD contributions to the
cross sections $\sigma^{\pm}_{LO}$. By using our program with the
same conditions as used in Table 3.1 of Ref.\cite{Wein}, we
calculated the LO cross section and the top-quark charge asymmetry
for the process $p\bar p \to t\bar t +jet+X$ at the Tevatron. We
obtained $\sigma_{LO}=1.582(2)~pb$ and $A_{FB,LO}^t=-7.70(6)\%$ by
taking $p_{T,jet,cut}=20~GeV$ and $\mu=m_t$, which are coincident
with those in Table 3.1 of Ref.\cite{Wein}.

\par
In Table \ref{tab2} we present the results of LO and NLO cross
sections($\sigma_{LO}$ and $\sigma^{QCD}$) and LO and NLO
forward-backward charge asymmetries of the top quark($A_{FB,
LO}^{t}$ and $A_{FB,NLO}^{t}$) for the process $p\bar p \to t\bar
t\gamma+X$ at the Tevatron. There we set $\mu=m_t$,
$\theta_{\gamma,jet}^{cut}=3^\circ$, and the photon transverse
momentum cut $p_{T,cut}^{(\gamma)}=20~GeV$, $30~GeV$ and $40~GeV$
respectively. The numerical results in the table show the NLO QCD
correction reduces evidently the absolute values of LO asymmetry,
e.g., in the case of $p_{T,cut}^{(\gamma)}=20~GeV$, the LO asymmetry
$|A_{FB,LO}^{t}|=17.24\%$ is cut down to $|A_{FB,NLO}^{t}|=11.41\%$
by the NLO QCD corrections. The absolute values for both LO and NLO
asymmetries are quantitatively increased a little bit with the
growing of $p_{T,cut}^{(\gamma)}$ from $20~GeV$ to $40~GeV$.

\begin{table}
\begin{center}
\begin{tabular}{|c|c|c|c|c|}
\hline $p_{T,cut}^{(\gamma)}(GeV)$ & $\sigma_{LO}$(fb) &
$\sigma^{QCD}$(fb) &
$A_{FB,LO}^{t}(\%)$ & $A_{FB,NLO}^{t}(\%)$  \\
\hline   20    & 43.79(4)  & 42.80(6)   & -17.24(7)  & -11.41(8)      \\
\hline   30    & 27.88(2)  & 26.94(4)   & -18.21(8)  & -11.52(9)      \\
\hline   40    & 19.09(1)  & 18.26(2)   & -18.85(8)  & -11.90(9)      \\
\hline
\end{tabular}
\end{center}
\begin{center}
\begin{minipage}{15cm}
\caption{\label{tab2} The LO and NLO cross sections and the  LO and
NLO forward-backward charge asymmetries of the top quark at the
Tevatron with $\mu = m_t$, $\theta_{\gamma,jet}^{cut} =3^\circ$(in
the proton-antiproton center-of mass system) and
$p_{T,cut}^{(\gamma)}=20~GeV$, $30~GeV$, $40~GeV$, respectively. }
\end{minipage}
\end{center}
\end{table}

\section{Summary}
\par
In this paper we calculate the complete NLO QCD corrections to the
top-pair production associated with a photon at the LHC and
Tevatron Run II. We investigate the dependence of the LO and NLO
QCD corrected integrated cross sections on the
factorization/renormalization energy scale. We present also the
predictions for LO and NLO QCD corrected charge asymmetries of
top-quarks at the Tevatron, and the LO and NLO differential cross
sections at the LHC and Tevatron. We find from our numerical
results that the NLO QCD radiative corrections obviously modify
the LO charge asymmetry of top-quark, integrated and differential
cross sections. And the uncertainty of the LO cross section due to
the introduced unphysical energy scale $\mu$, is significantly
improved by including NLO QCD corrections. Our numerical results
show that by taking $\mu=m_t$ the K-factors of the NLO QCD
corrections at the LHC and Tevatron RUN II are $1.524$ and
$0.977$, respectively.

\vskip 10mm
\par
\noindent{\large\bf Acknowledgments:} This work was supported in
part by the National Natural Science Foundation of
China(No.10875112, No.10675110).

\end{document}